\documentclass[preprint,showpacs,preprintnumbers,superscriptaddress,amsmath,amssymb]{revtex4}

\newcommand{\LamS}{$\Lambda(1405)$ }
\newcommand{\SigS}{$\Sigma^{0}(1385)$ }

\usepackage{graphicx}
\usepackage{dcolumn}
\usepackage{bm}

\begin{document}

\title{ Photoproduction of ${\boldmath \Lambda(1405)}$
 and ${\boldmath{\Sigma^{0}(1385)}}$ on the proton
 at ${\boldmath{E_\gamma}}$ = 1.5-2.4 GeV}

\author{M.~Niiyama}
\altaffiliation{Present address:  The Institute of Physical and Chemical Research, Wako, Saitama
\affiliation{Department of Physics, Kyoto University, Kyoto~606-8502, Japan}
 351-0198, Japan} 
\author{H.~Fujimura}
\altaffiliation{Laboratory of Nuclear Science, Tohoku University,Sendai~982-0826, Japan}
\affiliation{Department of Physics, Kyoto University, Kyoto~606-8502, Japan}
\author{D.S.~Ahn}
\affiliation{Research Center for Nuclear Physics, Osaka University, Ibaraki~567-0047, Japan} 
\author{J.K.~Ahn}
\affiliation{Department of Physics, Pusan National University, Busan~609-735, Korea}
\author{S.~Ajimura}
\affiliation{Research Center for Nuclear Physics, Osaka University, Ibaraki~567-0047, Japan} 
\author{H.C.~Bhang}
\affiliation{School of Physics, Seoul National University, Seoul, 151-747, Korea}
\author{T.H.~Chang}
\affiliation{Department of Information Management, Ling-Tung University, Taiwan}
\author{W.C.~Chang}
\affiliation{Institute of Physics, Academia Sinica, Taipei~11529, Taiwan}
\author{J.Y.~Chen}
\affiliation{Department of Physics, National Sun Yat-Sen University, Kaohsiung 804, 
Taiwan}
\author{S.~Dat\'{e}}
\affiliation{Japan Synchrotron Radiation Research Institute, Mikazuki~679-5198, Japan} 
\author{S.~Fukui}
\affiliation{Department of Physics and Astrophysics, Nagoya University, Nagoya, Aichi~464-8602, Japan}
\author{H.~Funahashi}
\affiliation{Osaka Electro-Communication University, 18-8 Hatsucho, Neyagawa, Osaka~572-8530, Japan}
\author{K.~Hicks}
\address{Department of Physics and Astronomy, Ohio University, Athens, Ohio~45701, USA} 
\author{K.~Horie} 
\affiliation{Department of Physics, Osaka University, Toyonaka~560-0043, Japan} 
\author{T.~Hotta} 
\affiliation{Research Center for Nuclear Physics, Osaka University, Ibaraki~567-0047, Japan} 
\author{K.~Imai}
\affiliation{Department of Physics, Kyoto University, Kyoto~606-8502, Japan}
\author{T.~Ishikawa}
\affiliation{Laboratory of Nuclear Science, Tohoku University,Sendai~982-0826, Japan}
\author{Y.~Kato}
\affiliation{Research Center for Nuclear Physics, Osaka University, Ibaraki~567-0047, Japan} 
\author{K.~Kino}
\affiliation{Research Center for Nuclear Physics, Osaka University, Ibaraki~567-0047, Japan} 
\author{H.~Kohri}
\affiliation{Research Center for Nuclear Physics, Osaka University, Ibaraki~567-0047, Japan} 
\author{S. Makino}
\affiliation{Wakayama Medical University, Wakayama, Wakayama 641-8509, Japan}
\author{T.~Matsumura}
\affiliation{Department of Applied Physics, National Defense Academy, Yokosuka~239-8686, Japan}
\author{T.~Mibe}
\affiliation{High Energy Accelerator Reseach Organization, KEK, 1-1 Oho Tsukuba, Ibaraki~305-0801, Japan}
\author{K.~Miwa}
\affiliation{\it Department of Physics, Tohoku University, Sendai~980-0861, Japan}
\author{M.~Miyabe}
\affiliation{Department of Physics, Kyoto University, Kyoto~606-8502, Japan}
\author{N.~Muramatsu}
\affiliation{Research Center for Nuclear Physics, Osaka University, Ibaraki~567-0047, Japan} 
\author{M.~Nakamura}
\affiliation{Wakayama Medical University, Wakayama, Wakayama 641-8509, Japan}
\author{T.~Nakano}
\affiliation{Research Center for Nuclear Physics, Osaka University, Ibaraki~567-0047, Japan} 
\author{Y.~Nakatsugawa}
\affiliation{Department of Physics, Kyoto University, Kyoto~606-8502, Japan}
\author{Y.~Ohashi}
\affiliation{Japan Synchrotron Radiation Research Institute, Mikazuki~679-5198, Japan} 
\author{D.S.~Oshuev}
\affiliation{Institute of Physics, Academia Sinica, Taipei~11529, Taiwan}
\author{N.~Saito}
\affiliation{High Energy Accelerator Reseach Organization, KEK, 1-1 Oho Tsukuba, Ibaraki~305-0801, Japan}
\author{T.~Sawada}
\affiliation{Research Center for Nuclear Physics, Osaka University, Ibaraki~567-0047, Japan} 
\author{Y.~Sugaya}
\affiliation{Department of Physics, Osaka University, Toyonaka~560-0043, Japan} 
\author{M.~Sumihama}
\affiliation{Research Center for Nuclear Physics, Osaka University, Ibaraki~567-0047, Japan} 
\author{J.L.~Tang}
\affiliation{ Department of Physics, National Chung Cheng University, Taiwan} 
\author{M.~Uchida}
\affiliation{Department of Physics, Tokyo Institute of Technology, Tokyo 152-8551, Japan} 
\author{C.W.~Wang}
\affiliation{Institute of Physics, Academia Sinica, Taipei~11529, Taiwan}
\author{T.~Yorita}
\affiliation{Research Center for Nuclear Physics, Osaka University, Ibaraki~567-0047, Japan} 
\author{M.~Yosoi}
\affiliation{Research Center for Nuclear Physics, Osaka University, Ibaraki~567-0047, Japan}


\date{\today}

\begin{abstract}
 Differential cross sections for $\gamma p \rightarrow
 K^+\Lambda(1405)$ and $\gamma p \rightarrow K^+\Sigma^0(1385)$ 
 reactions have been measured in the photon energy range from 1.5 to 2.4 GeV
 and the angular range of $0.8<\cos(\Theta)<1.0$ for the $K^+$ scattering
 angle in the center-of-mass system.
 This data is the first measurement of the $\Lambda(1405)$ photoproduction
 cross section. 
 The lineshapes of \LamS measured in $\Sigma^+\pi^-$ and $\Sigma^-\pi^+$ 
 decay modes were different with each other, indicating
 a strong interference of the isospin 0 and 1 terms of the $\Sigma\pi$
 scattering amplitudes.
 The ratios of \LamS production to \SigS production were measured in two
 photon energy ranges: near the production threshold ($1.5<E_\gamma<2.0$ GeV)
 and far from it ($2.0 <E_\gamma<2.4$ GeV).
 The observed ratio decreased in the higher photon energy region, which
 may suggest different production mechanisms and internal structures
 for these hyperon resonances.
 
\end{abstract}



\maketitle

\section{Introduction}

 Recently, the photoproduction of $\Lambda$ and $\Sigma^0$  hyperons
 off protons has been studied with high statistics
 data~\cite{Tran1998,McNabb2004,Glander2004,Sumihama2006}, mainly motivated
 by missing resonances which could couple to the $KY$ channels
 \cite{Capstick1998}.
 The authors of Ref~\cite{Capstick1998} also predicted nucleon resonances which
 couple to the $KY^*$ channels, $K\Sigma(1385)$ and $K\Lambda(1405)$.
 In addition to searching for missing resonances, it is of interest to
 study the internal structure of hyperon resonances. 
 The \SigS hyperon is firmly established as a $q^3$ baryon.
 On the contrary, the internal structure of the \LamS hyperon is ambiguous.
 In the quark model, \LamS is assigned as a $p$-wave $q^3$ baryon~\cite{Isgur1978}.
 However, it is also widely discussed as a candidate for
 a meson-baryon molecular state~\cite{Dalitz1967,Kaiser1995,Nacher1999} or
 a $q^4\bar{q}$ pentaquark baryon~\cite{Inoue2007}.
 An overview of the study of the \LamS before the year 1998 can be found in
 Ref.~\cite{Dalitz2000}.
 Recent lattice calculations also reported a non-$q^3$ structure for the
 \LamS~\cite{Nemoto2003,Ishii2007}. 
 Based on the meson-baryon molecular picture of the $\Lambda(1405)$,
 a model calculation using the chiral Lagrangian and unitary coupled channel
 approach succeeded to dynamically generate the mass spectrum of
 $\Lambda(1405)$~\cite{Nacher1999}.
 The same model also predicted a second pole for the \LamS with a
 mass of 1.425 GeV/c$^2$, which strongly couples to $\bar{K}N$~\cite{Jido2005},
 and experimental evidence for this second pole has been 
 reported~\cite{Prakhov2004,Magas2005}.
 It is important for the study of the internal structure of the \LamS 
 to measure the pole position and width of this hyperon resonance from 
 the invariant mass distribution.
 In order to extract this information using the $\Sigma\pi$ final state,
 one has to take into account the effect of the interference between
 the isospin 0 and 1 amplitudes of the
 $\Sigma\pi$ interaction~\cite{Nacher1999} and the interference between
 the s-wave and p-wave amplitudes~\cite{JidoPrivate}.

 The cross section of hyperon photoproduction depends on its production
 mechanism and the form factor of the hyperon. 
 Photoproduction of \SigS has been measured by the CLAS Collaboration
 at TJNAF with a photon energy range of 1.5-3.8 GeV~\cite{Guo2006} and
 by old bubble chamber experiments~\cite{Cambridge1967,Erbe1969}.
 A theoretical calculation by Oh {\it et al.}~\cite{Oh2007}
 using an effective Lagrangian was then compared with the preliminary data of
 total cross section by the CLAS experiment, and the contributions from
 nucleon resonances were discussed.
 Experimentally, the \LamS has been studied in meson-induced and
 proton-induced reactions so
 far~\cite{Prakhov2004,Thomas1973,Hemingway1985,Zychor2008}.
 However, understanding of the photoproduction of \LamS is very limited
 because of the lack of experimental data.
 Theoretically, Nacher {\it et al.}~\cite{Nacher1999} predicted the cross
 section of \LamS photoproduction to be 5~$\mu$b/GeV at the peak of the
 invariant mass spectrum of \LamS using the chiral unitary model.
 Lutz and Soyeur calculated the differential cross section for
 the sum of $\Sigma(1385)$ and $\Lambda(1405)$ photoproduction using a chiral 
 coupled-channel effective model~\cite{Lutz2005}. 
 In both theoretical calculations, the effect of the interference with the p-wave
 amplitude was neglected.
 Experimentally, the contribution of the p-wave amplitude can be studied 
 by increasing the photon energy from the production threshold.
  In this paper, the differential cross section of \LamS photoproduction
 and its comparison with that of \SigS are reported for the first time. 
 The cross section was measured in two photon energy bins: 
 near production threshold region and far from it.

\section{Experimental procedure and setup}
 
 The experiment was carried out at the Laser-Electron Photon facility 
 at SPring-8 (LEPS) \cite{Sumihama2006,Nakano2001}.
 The data were collected using two different experimental setups,
 referred to hereafter as data set (I) and data set (II).
 Photoproduction of $\Lambda$ and $\Sigma^0$ has been
 studied at LEPS using high statistics data with a liquid-hydrogen
 target (data set (I))~\cite{Sumihama2006}.
 The statistics of this data set were sufficiently high to determine the
 differential cross sections accurately for the ground state $\Lambda$ and
 $\Sigma^0$ hyperons. Nevertheless, in principle, it is impossible to
 separate \LamS and \SigS from a missing mass of
 $\gamma p \rightarrow K^+ Y^*$ reactions, $MM(K^+)$,
 because the intrinsic widths of these resonances,
 36 MeV/c$^2$ for \SigS and 50 MeV/c$^2$ for $\Lambda(1405)$, are much larger
 than their mass difference.
 In order to distinguish these two, a time projection chamber (TPC) was
 used for data set (II) together with
 the LEPS spectrometer to facilitate the detection of the decay products
 of these hyperon resonances. Using this data set, the production ratio
 between \SigS and \LamS
 was fixed from the yields of \SigS and \LamS measured from their
 $\Lambda \pi^0$ and $\Sigma^\pm\pi^\mp$ decay modes, respectively.
 The absolute values of the differential cross sections were then
 obtained from the $MM(K^+)$ distribution measured in data set (I)
 with the input of
 the production ratio between \SigS and $\Lambda(1405)$ in the common
 detector acceptance of the two data sets.


 At LEPS, photons with a maximum energy of 2.4 GeV are produced by Compton
 back-scattering of laser photons with a wavelength of 351 nm from
 the 8-GeV electrons in the SPring-8 storage ring. 
 To measure the energy of each photon, the recoiling electrons
 were momentum analyzed by a bending magnet of the storage
 ring and detected by a tagging counter (tagger) inside the ring. 

\begin{figure}[hbtp]
\begin{center}
 \includegraphics[width=10cm,height=8cm,keepaspectratio]
{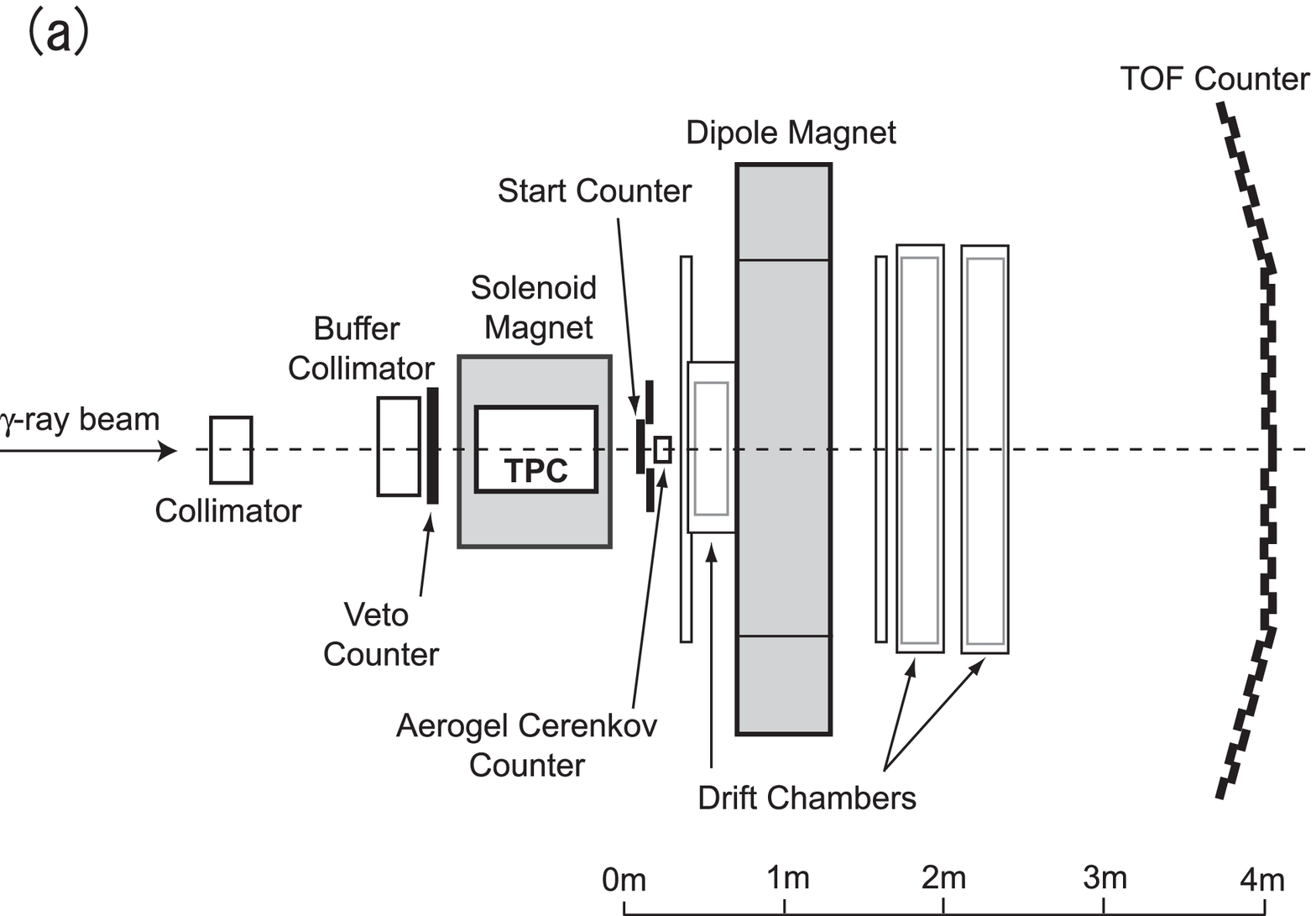}
\end{center}
\vspace{1.0cm}
\begin{minipage}{0.4\hsize}
\begin{center}
 \includegraphics[width=6.0cm,height=6.0cm,keepaspectratio]
{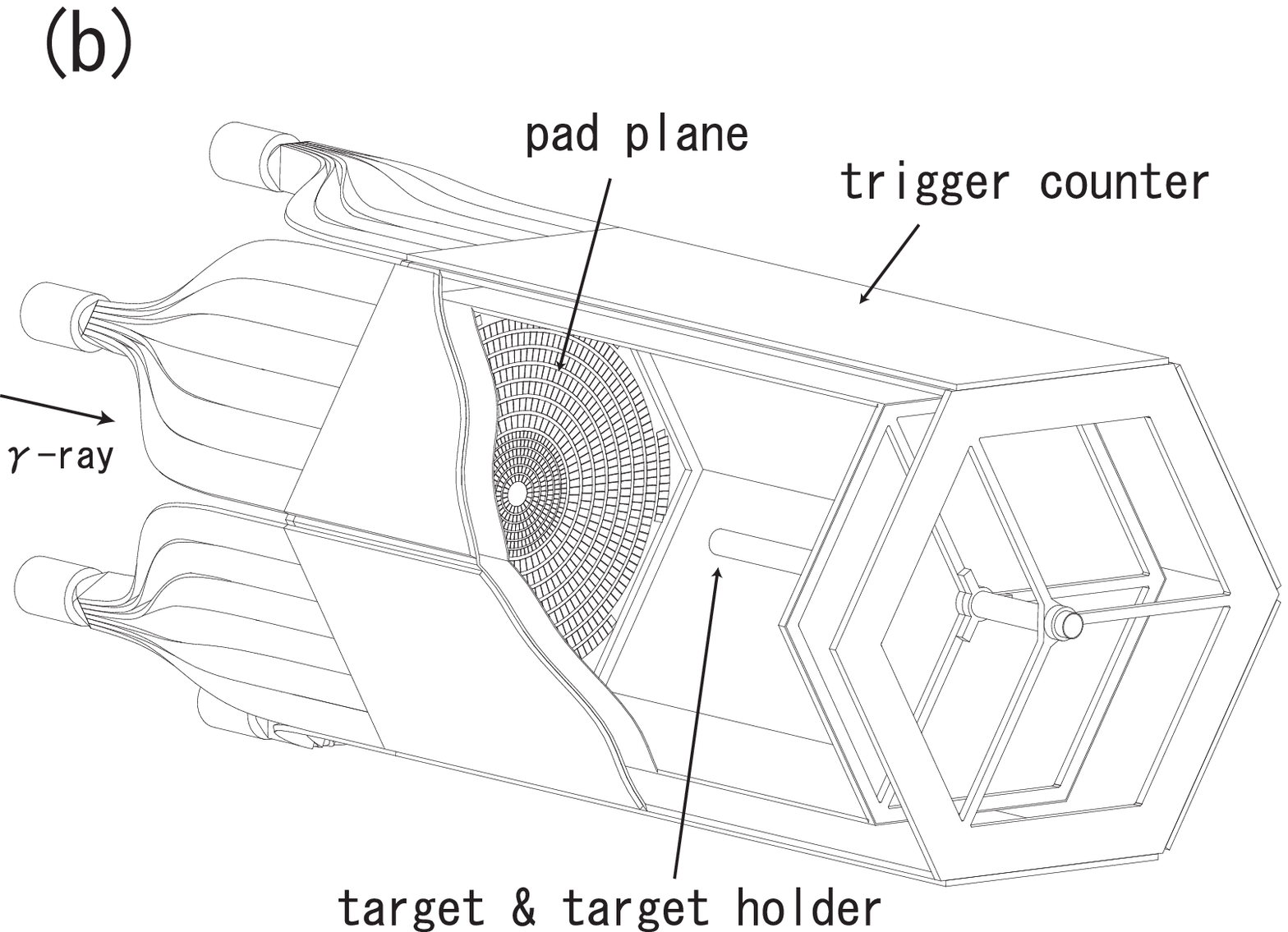}
\end{center}
\end{minipage}
\vspace{1.0cm}
\begin{minipage}{0.4\hsize}
\begin{center}
 \includegraphics[width=5.5cm,height=5.5cm,keepaspectratio]
{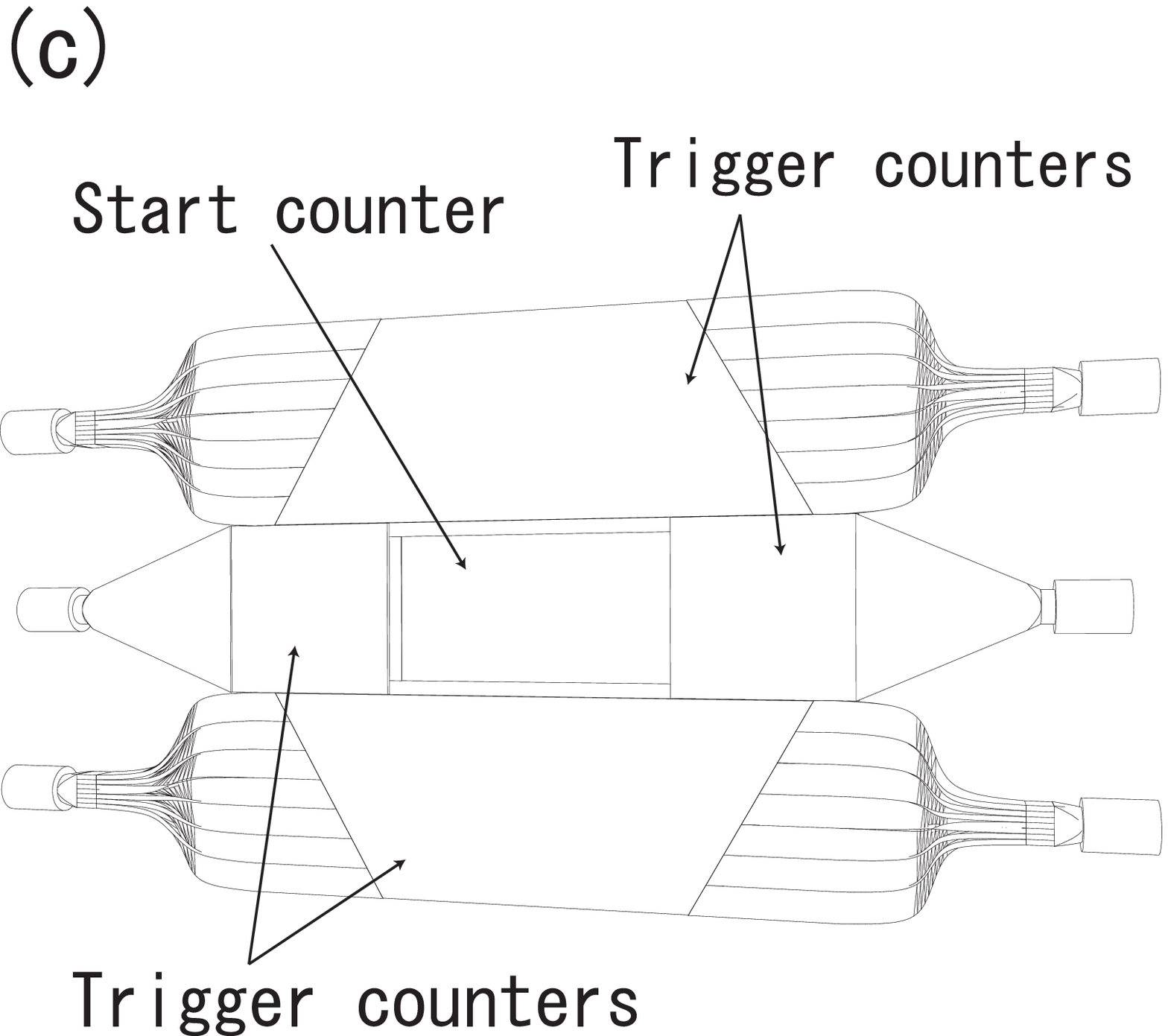}
\end{center}
\end{minipage}
\caption{(a) Schematic top view of experimental apparatus.
 (b) Schematic view of the TPC and trigger counters. (c) Schematic view of
the trigger counters placed between the TPC and the LEPS spectrometer.}
\label{lepsSetup}
\end{figure}

 Forward going $K^+$'s from the $\gamma p \rightarrow K^+ X$ reaction
 were detected in the LEPS spectrometer, which consisted of a dipole
 magnet, a silicon-strip vertex detector, three multiwire drift chambers,
 a start counter (SC) just downstream of the target, and a time-of-flight (TOF)
 hodoscope placed downstream of the tracking detectors.
 Electron-positron pairs were vetoed by an aerogel \v{C}erenkov
 detector just after the SC.
 Figure~\ref{lepsSetup} (a) shows a schematic drawing of the LEPS spectrometer.
 More details of the LEPS spectrometer can be found in Ref.~\cite{Sumihama2006}.

 The TPC, shown in Fig.~\ref{lepsSetup} (b), was used for data set (II) to
 detect the decay topology of low momentum hyperons, $\Lambda$ and $\Sigma^\pm$,
 originating from the decay of $Y^*$ hyperon resonances.
 Cylindrical targets of CH$_2$ and carbon with a diameter of 24~mm and
 lengths of 47~mm (CH$_2$) and 22~mm (carbon) were placed inside the TPC.
 The TPC had an active volume of cylindrical shape with a radius of 200 mm
 and a length of 770 mm, which was filled with P10 gas
 (Ar : CH$_4$ 90\% : 10\%). 
 The TPC was immersed in a solenoidal magnetic field of 2 Tesla. 
 The signals from the TPC were read through 1055 cathode pads which were arranged
 in 14 circular rows and were amplified by preamplifiers and 
 shaping amplifiers. The wave form of each signal was digitized by flash ADC
 with 10 bit resolution and a 40 MHz sampling rate.
 The typical spatial resolutions were found using cosmic ray tracks to be
 $\sim 400$ $\mu$m in pad plane and $\sim 700$ $\mu$m in the z direction.
 Details on the TPC can be found in Ref.~\cite{Niiyama2003}.

 The TPC was surrounded by ten scintillation counters to
 detect charged particles passing through the active volume of the TPC. 
 Six scintillation counters were equipped on the sides of the TPC,
 and the remaining four were placed between the TPC and the LEPS
 spectrometer (Fig.~\ref{lepsSetup} (c)).
 The trigger used for this analysis was a coincidence between the tagger,
 the SC, the TOF wall, and any two of the ten scintillation counters surrounding
 the TPC.
 The aerogel \v{C}erenkov veto counter was also include at the trigger level.

\section{Analysis}

 First, the analysis for the production ratio between \SigS and \LamS using
 data set (II) is described, followed by the determination of the absolute
 values of the differential cross sections using data set (I).
 The ratio of the production cross section between \SigS and \LamS was obtained
 from the following two reactions:
 \begin{eqnarray}
  \gamma p &\rightarrow& K^+ \Sigma^{0}(1385)\rightarrow K^+ \Lambda \pi^0\rightarrow K^+ p \pi^- \pi^0, \\
\mathrm{and} \qquad  \gamma p &\rightarrow& K^+ \Lambda(1405)\rightarrow K^+ \Sigma^\pm \pi^\mp\rightarrow K^+ n \pi^+ \pi^- .
 \end{eqnarray}

 The yield of \SigS production was measured from reaction (1) because
 \LamS is prohibited from decaying into $\Lambda \pi^0$ by isospin conservation.
 Since the detectors were not sensitive to photons,  $\pi^0$'s were measured through
 the missing mass technique.
 The contamination of \SigS in reaction (2) was estimated from the yield measured
 in reaction (1). The yield of \LamS production was extracted from 
 reaction (2) after subtracting the contamination of \SigS production.
 The production ratio was obtained after taking into account 
 detector acceptance and the decay branches of these hyperons.
 The differential cross sections of \SigS and \LamS production were
 then obtained in two photon energy bins: $1.5<E_\gamma<2.0$ GeV and 
 $2.0<E_\gamma<2.4$ GeV.



 For this analysis, events of the type $\gamma p \rightarrow K^+ X$  were
 selected by identifying a $K^+$ in the LEPS spectrometer.
 After reconstructing the mass of the each track from the momentum
  and TOF information, a 4$\sigma$ cut was used to select the $K^+$'s
  taking into account the momentum-dependence of the mass
 resolution~ \cite{Sumihama2006}.
 To reject kaons which decayed within the spectrometer, events were
 required to have a $\chi^2$ probability for track fitting of
 greater than 0.02.
 Events produced at the nuclear target were selected by their closest points
 between the $K^+$ track and the beam axis.
 For the photon energy measurement, we required a single electron track be
 reconstructed by the tagging counter.
 The numbers of events surviving the $K^+$ selection cuts are summarized in
 Table~\ref{cutSummary_K}.

{\small
\begin{table}[tbp]
\caption{Number of events after selection cuts.}
\label{cutSummary_K}
\begin{tabular}{l r r}
\hline 
Cut by the LEPS spectrometer & events (CH$_2$) & events (C) \\ 
$K^+$ selection & 1.39$\times 10^5$ & 7.38$\times 10^4$  \\ 
$K^+$ decay-in-flight rejection & 1.30$\times 10^5$ & 6.88$\times 10^4$  \\ 
Vertex selection at the target & 1.26$\times 10^5$ & 6.68$\times 10^4$  \\ 
One recoil electron in the tagging counter & 1.12$\times 10^5$ & 5.92$\times 10^4$  \\ 
\hline 
Cut by the TPC for \SigS  & events (CH$_2$) & events (C) \\ 
Proton and $\pi^-$ identification by TPC & $5.45\times 10^4$ & $2.81\times 10^4$  \\ 
$\Lambda$ selection  & $2.69\times 10^4$ & $1.46\times 10^4$  \\ 
\hline 
Cut by the TPC for \LamS  &  events (CH$_2$) &  events (C) \\ 
$\pi^+\,\pi^-$ selection in TPC & $4.92 \times 10^4$ & $2.46\times 10^4$  \\ 
$MM^2(K^+ p \pi^-) < -0.05$ (GeV/c$^2$)$^2$ & $3.33\times 10^4$ & $1.73\times 10^4$  \\ 
neutron selection & $6.46\times 10^3$ & $2.59\times 10^3$  \\ 
$\chi^2$ probability cut of C2-fit & $2.98\times 10^3$ & $7.96\times 10^2$  \\ 
\hline 
\end{tabular} 
\end{table}
}

To study the \SigS production, we required the following cut conditions:
(i) selection of a $K^+$ in the LEPS spectrometer, 
(ii) identification of a proton and a $\pi^-$ in the TPC,
(iii) selection of those events for which the mass of
 a $(p \, \pi^-)$ pair corresponds to that of the $\Lambda$.
 The numbers of events which survived these cuts are shown in 
 Table~\ref{cutSummary_K}.
 To match the acceptance of the LEPS spectrometer between data sets
 (I) and (II), the scattering angle of the $K^+$ in the center-of-mass
 frame, $\Theta_{K_{CM}}$, was required to be
 $0.8<\cos(\Theta_{K_{CM}})<1.0$.

\begin{figure}[htbp]
\begin{tabular}{c c}
\begin{minipage}{0.5\hsize}
\begin{center}
 \includegraphics[width=7.0cm,height=7.0cm,keepaspectratio]
{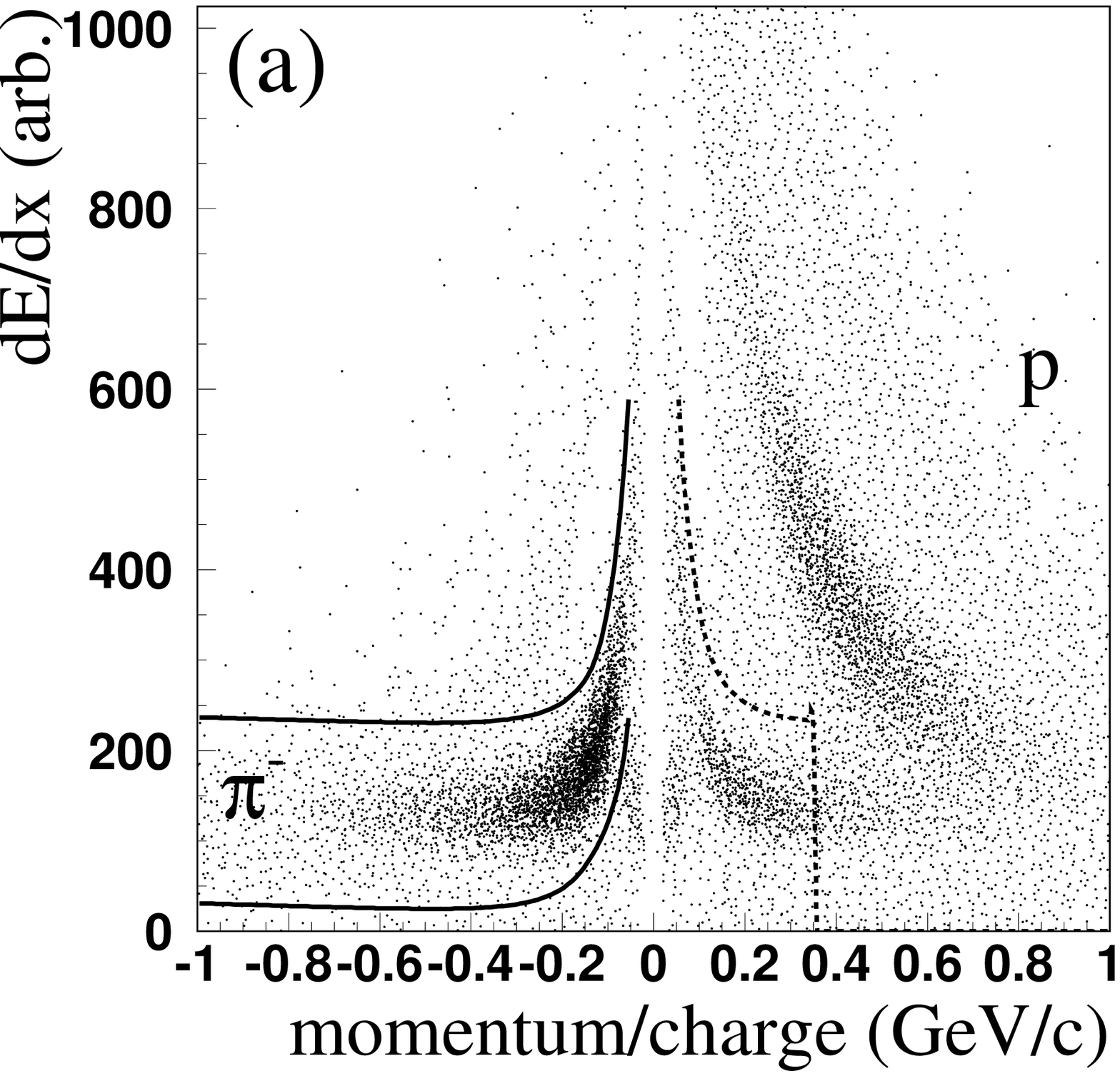}
\end{center}
\end{minipage} & %
\begin{minipage}{0.5\hsize}
\begin{center}
 \includegraphics[width=7.0cm,height=7.0cm,keepaspectratio]
{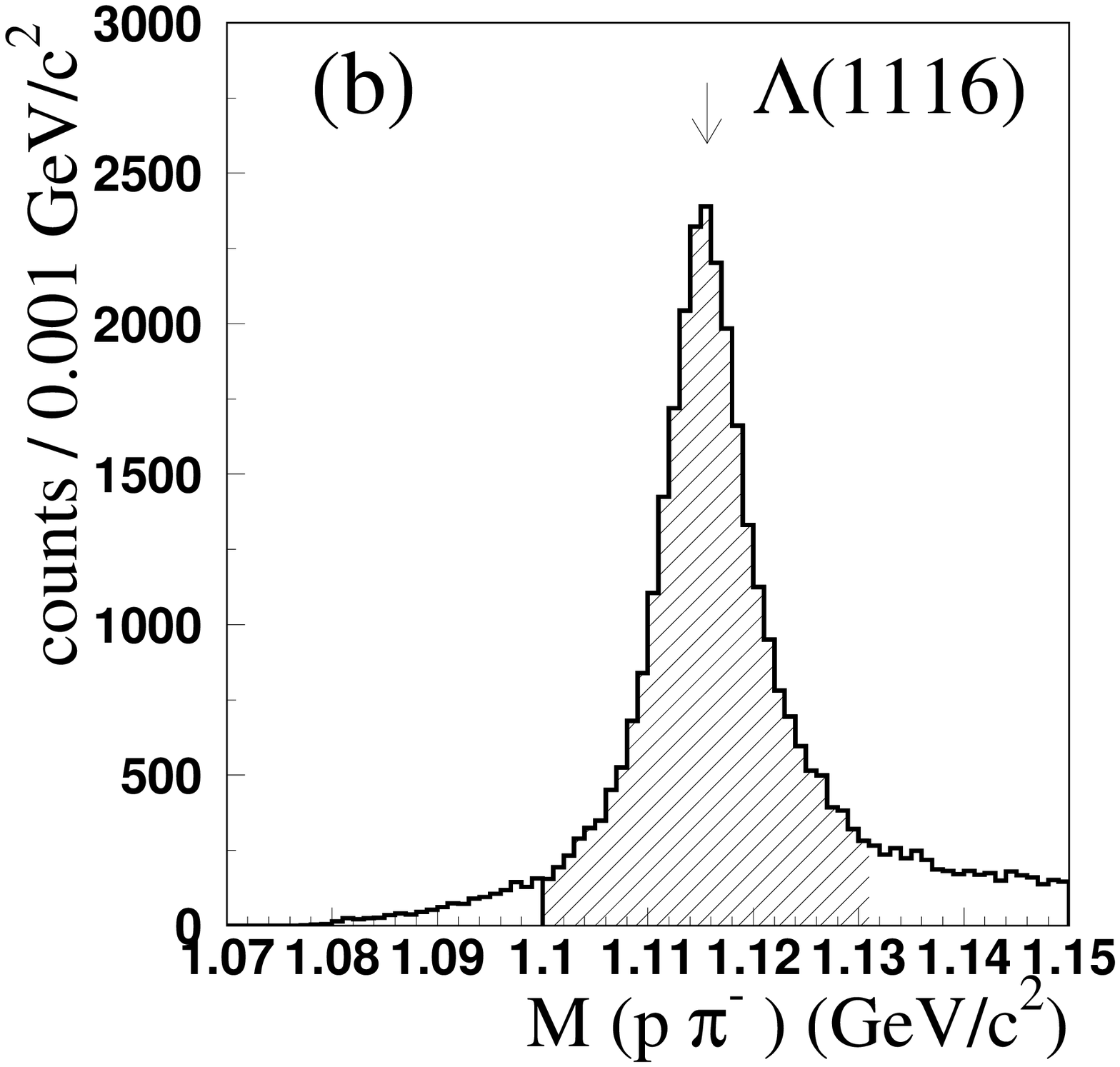}
\end{center}
\end{minipage} 
\end{tabular}
\caption{
 (a) Correlation plot of the energy deposition
 and momentum/charge for charged particles measured by the TPC.
 Solid lines and dashed line show boundaries for $\pi^-$ and proton selection,
 respectively.
 (b) The invariant mass spectrum of  ($p \, \pi^-$) pairs.
}
\label{PID_Lambda}
\end{figure}

 The candidate tracks of a $(p\, \pi^-)$ pair were identified from the
 truncated mean of the energy deposition ($dE/dx$) measured by the pads of the TPC.
 Fig.~\ref{PID_Lambda} (a) shows
 a correlation plot between $dE/dx$ and momenta of charged particles
 detected in the TPC
 for events where a $K^+$ was found in the LEPS spectrometer.
 Protons and $\pi^-$'s were selected by the dashed and solid curves, respectively.
  Fig.~\ref{PID_Lambda} (b) shows the invariant mass
  spectrum of $(p \, \pi^-)$ pairs, $M(p\pi^-)$. 
  The mass and width ($\sigma$) were obtained from a Gaussian fit to the data 
  to be 1115.4 $\pm$ 0.4 MeV/c$^2$ and 3.9 $\pm$ 0.5 MeV/c$^2$, respectively,
  and are consistent with the value of the mass listed in the PDG 
  and the expected width of 4.0 MeV/c$^2$
  determined by a Monte Carlo (MC) simulation.
  The events in the hatched area were retained for
  the analysis of \SigS production.

\begin{figure}[htbp]
\begin{tabular}{c c}
\begin{minipage}{0.5\hsize}
\begin{center}
 \includegraphics[width=7.0cm,height=7.0cm,keepaspectratio]
{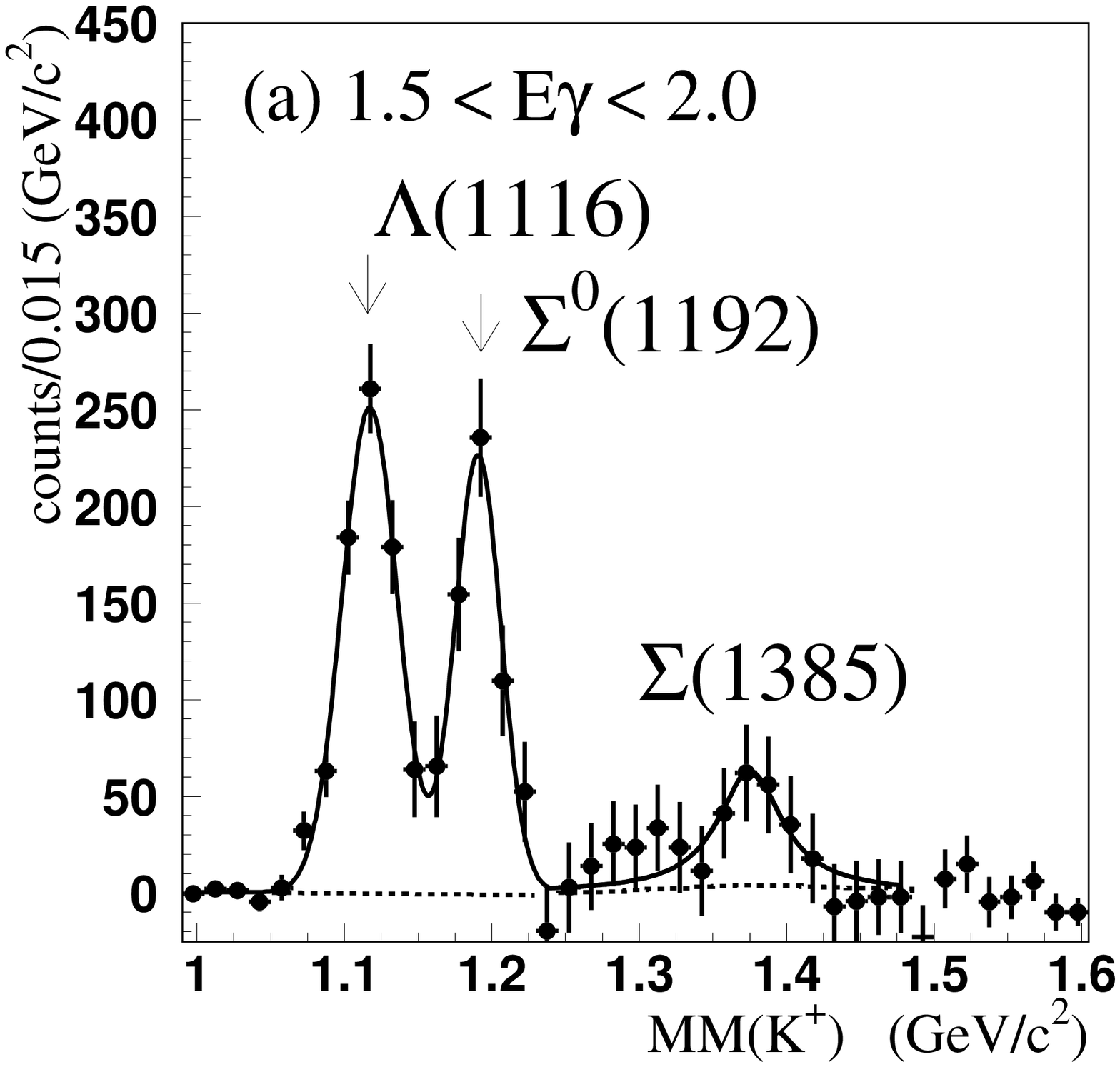}
\end{center}
\end{minipage} & %
\begin{minipage}{0.5\hsize}
\begin{center}
 \includegraphics[width=7.0cm,height=7.0cm,keepaspectratio]
{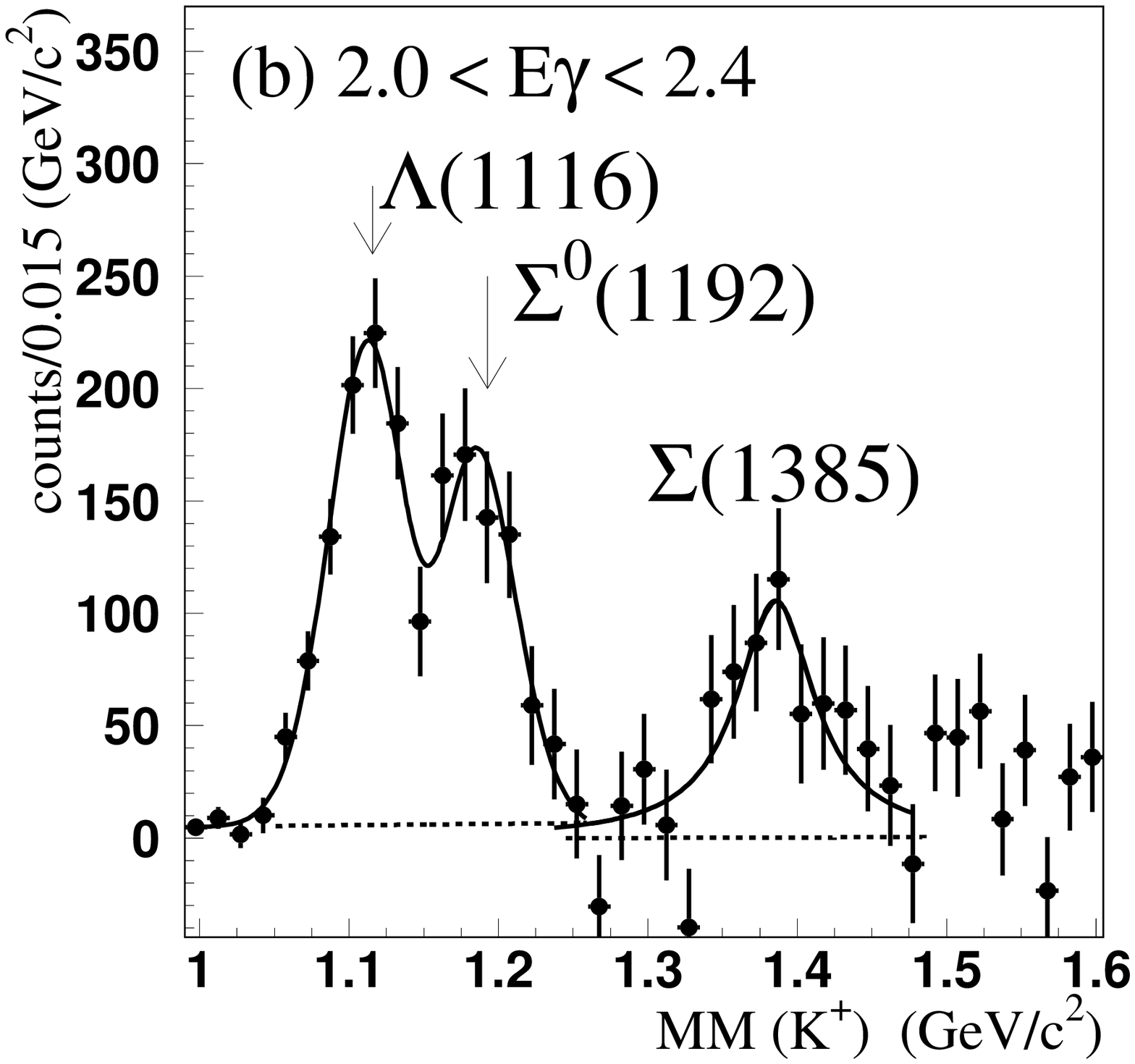}
\end{center}
\end{minipage} \\
\end{tabular}
\caption{
 $MM(K^+)$ distribution after the $\Lambda$ selection cut for
 two photon energy bins: (a) $1.5<E_\gamma<2.0$ GeV and (b) $2.0<E_\gamma<2.4$ GeV,
 respectively.
}
\label{S1385}
\end{figure}

  First, the yield of \SigS was extracted from the peak in the missing mass
  spectrum of the $\gamma p \rightarrow K^+ X$ reaction for events 
  which the mass of a $(p\,\pi^-)$ pair corresponded to that of 
  the $\Lambda$. 
  The spectrum for free protons was obtained from the spectrum
  for CH$_2$ by subtracting the one for carbon after normalizing each spectrum
  by the number of photons and the number of
  carbon nuclei in each target. 
  Before the subtraction, the spectrum for carbon was smoothed by smearing
  the photon energies with the experimental resolution in order to reduce
  the effect of statistical fluctuations.
  For each event, the photon energy was smeared by Gaussian random numbers
  with a width of 15 MeV, and the missing masses were calculated using the
  smeared photon energies.  
  Fig.~\ref{S1385} (a) and (b) show the $MM(K^+)$ spectrum for free protons
  in CH$_2$ for two photon energy bins:
  $1.5<E_\gamma<2.0$ GeV and $2.0<E_\gamma<2.4$ GeV, respectively.
  From these spectra, the masses of the $\Lambda(1116)$ and $\Sigma^0(1192)$ were
  obtained as $1115 \pm 2$ MeV/c$^2$ and $1192 \pm 3$ MeV/c$^2$,
  respectively, in agreement with the PDG values.
  The observed widths (RMS) of $\Lambda(1116)$ were $19 \pm 2$ MeV/c$^2$ 
  and $27 \pm 2$ MeV/c$^2$ for $1.5<E_\gamma<2.0$~GeV and $2.0<E_\gamma<2.4$~GeV,
  respectively, which are consistent with the expected values of 19 MeV/c$^2$ 
  and 24~MeV/c$^2$ for each photon energy bin.
  Those of $\Sigma(1192)$ were also consistent with expected values.
  For the resonance around 1.4 GeV/c$^2$, we assumed a Breit-Wigner shape, 
  neglecting any distortion due to the small contamination of $\Lambda(1405)$.
  The effect of this contamination is considered below.
  The mass of the \SigS was found from a Breit-Wigner fit with a linear
  background assumption to be $1375 \pm 10 $~MeV/c$^2$ and $1386 \pm 7$~MeV/c$^2$
  for $1.5<E_\gamma<2.0$~GeV and $2.0<E_\gamma<2.4$~GeV, respectively.
  The observed widths (FWHM) were $53 \pm 29$ MeV/c$^2$ and $63 \pm 24$ MeV/c$^2$,
  which are consistent with the expected width of 50 MeV/c$^2$.
  
  In order to obtain the yield of $\Sigma^{0}(1385)$, two background contributions
  were investigated: 

\begin{eqnarray}
\gamma p &\rightarrow& K^+ \Lambda(1405) \rightarrow  K^+\Sigma^0 \, \pi^0 \rightarrow K^+\Lambda \gamma \, \pi^0  \rightarrow K^+p\, \pi^- \, \gamma \, \pi^0, \\
\mathrm{and} \qquad \gamma p &\rightarrow& K^+ \Lambda(1405) / \Sigma^{0}(1385) \rightarrow  K^+ \Sigma^+ \, \pi^- \rightarrow K^+ p \,\pi^0 \,\pi^- .
\end{eqnarray}

\begin{figure}[htbp]
\begin{tabular}{c c}
\begin{minipage}{0.5\hsize}
\begin{center}
 \includegraphics[width=7.0cm,height=7.0cm,keepaspectratio]
{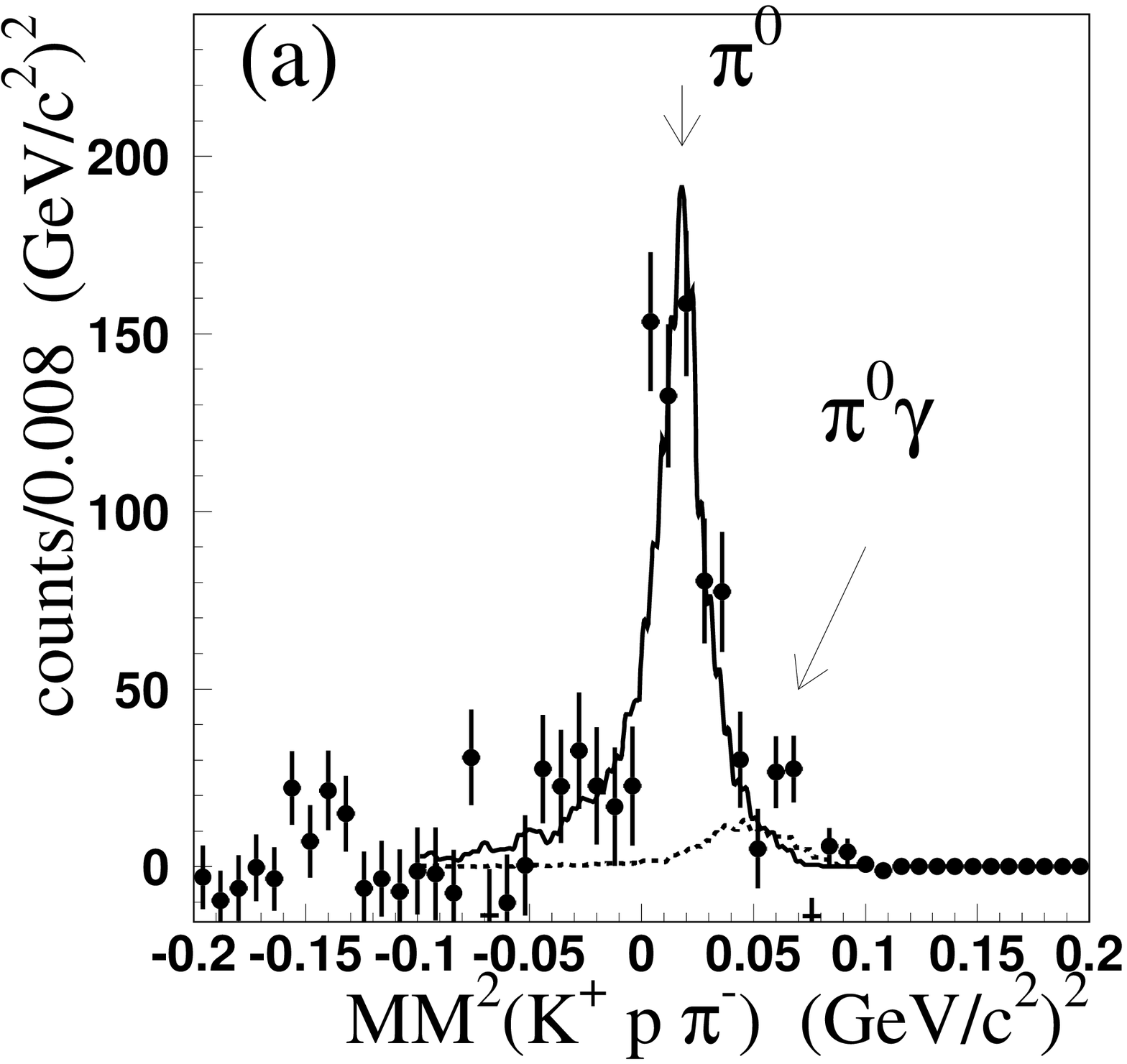}
\end{center}
\end{minipage} & %
\begin{minipage}{0.5\hsize}
\begin{center}
 \includegraphics[width=7.0cm,height=7.0cm,keepaspectratio]
{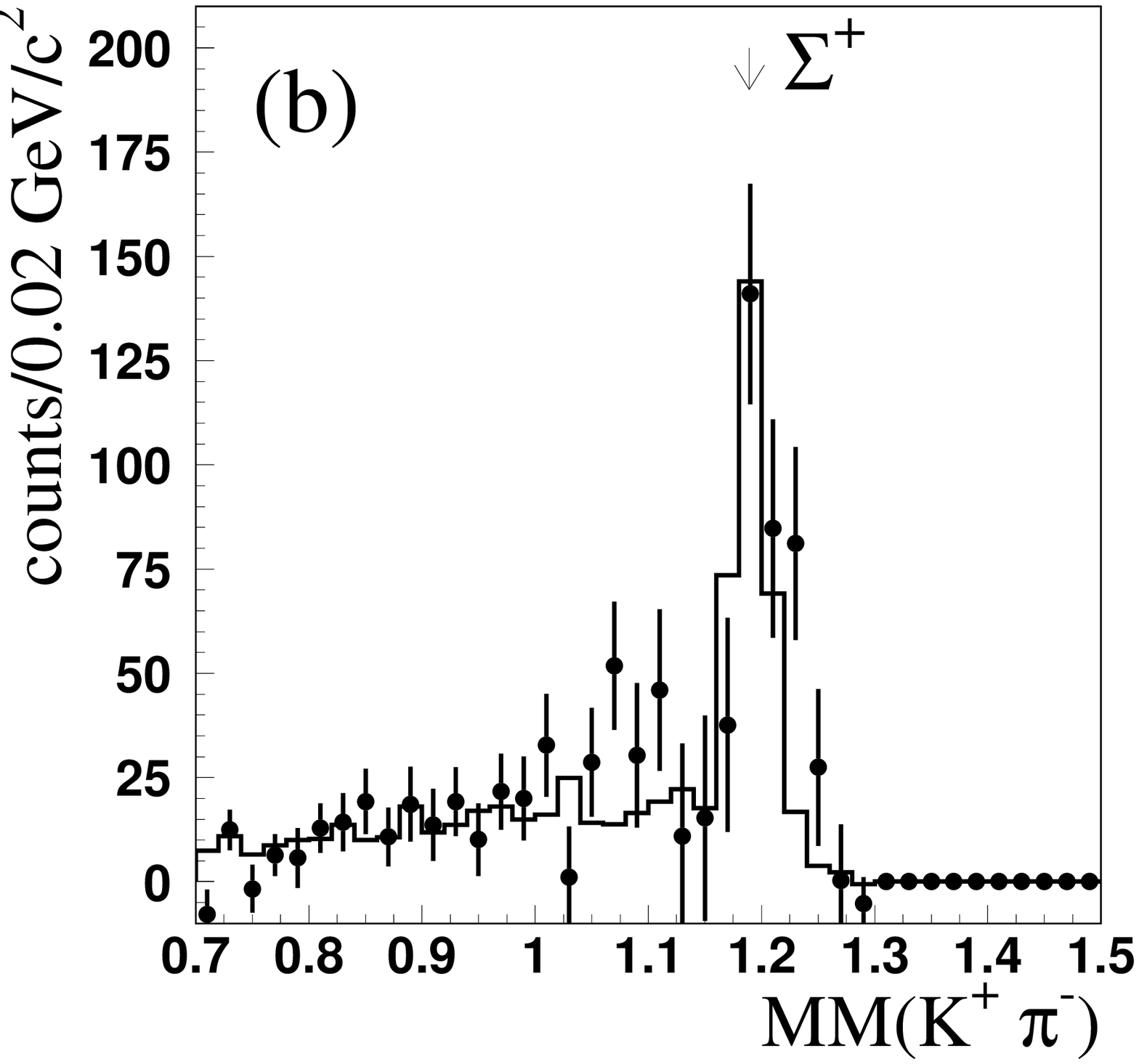}
\end{center}
\end{minipage} \\
\end{tabular}
\caption{
 (a) The spectrum of $MM^2(K^+ \, p \, \pi^-)$.
 The solid and hatched histograms are
 the expected spectra for \SigS and \LamS photoproduction, respectively,
 as determined by the MC simulation.
(b) $MM(K^+ \pi^-)$ distribution after the $\Lambda$ rejection cut and
 \LamS selection cut, $1.3<MM(K^+)<1.45$ GeV/c$^2$.
}
\label{S1385-bk}
\end{figure}

 The background from reaction (3) was estimated from the
 missing mass of the $\gamma p \rightarrow K^+ p \pi^- X$ reaction,
 $MM(K^+p\pi^-)$, where
 $X = \pi^0$ for \SigS production and  $X = \pi^0 \gamma$ for \LamS production.
 Fig.~\ref{S1385-bk} (a) shows the distribution of missing mass squared, 
 $MM^2(K^+ p \pi^-) $, for free protons in the CH$_2$ target
 for events which passed the $\Lambda$ selection cut and a
 \LamS/\SigS selection cut of $1.3 < MM(K^+) < 1.45$ GeV/c$^2$. 
 The peak at 0.018 (GeV/c$^2$)$^2$ corresponds to the square of the $\pi^0$ mass. 
 The solid and hatched histograms show the expected spectra 
 determined from MC simulation for \SigS production and for the background
 reaction (3), respectively.
 The normalization factors for these spectra were determined by fitting to the data.
 The contamination from reaction (3) was found to be 8\% of 
 the number of events in the \SigS mass region.

 The remaining contamination from reaction (4) was estimated 
 from the yield of $\Sigma^+$ in the missing
 mass of the $\gamma p \rightarrow K^+\pi^-X$ reaction, $MM(K^+\pi^-)$,
 for events in the \SigS mass region which were rejected by the
 $\Lambda$ selection cuts. The resulting spectrum for $MM(K^+\pi^-)$ is
 shown in Fig.~\ref{S1385-bk} (b).
 The solid curve shows the spectrum of $MM(K^+\pi^-)$ generated by a
 MC simulation of the background reaction (4) and events
 caused by a proton or a $\pi^+$ misidentified as a $K^+$ by the
 LEPS spectrometer.
 The acceptance of the $\Lambda$ rejection cut was calculated using the MC
 simulation, and the contribution from reaction (4) was found
 to be about 12\% of the number of events in \SigS mass region.
  The yield of \SigS was obtained by subtracting
  the contamination from reactions (3) and (4) and was found to be
  $255 \pm 55$ events and $525\pm111$ events
  for $1.5<E_\gamma<2.0$ GeV and $2.0<E_\gamma<2.4$ GeV, respectively.

 Next, the production of \LamS was measured using the
 $\gamma p \rightarrow K^+ \Lambda(1405) \rightarrow K^+ \Sigma^\pm\pi^\mp
 \rightarrow K^+ \pi^+ \pi^- n$ reaction. 
 The following cut conditions were imposed:
 (i)   selection of a $K^+$ in the LEPS spectrometer,
 (ii) identification of a $\pi^+$ and a $\pi^-$ in the TPC,
 (iii) rejection of events where a proton was misidentified
 as a $\pi^+$ in the TPC, 
 (iv)  selection of a neutron in the missing mass of the
       $\gamma p \rightarrow K^+ \pi^+ \pi^- X$ reaction, and
 (v)  selection of events with a good $\chi^2$ probability of kinematic fit
 assuming neutron and $\Sigma^\pm$ masses.
 The numbers of events remaining after these cuts are shown in Table~\ref{cutSummary_K}.
 
\begin{figure}[htbp]
 \begin{tabular}{c c}
  \begin{minipage}{0.5\hsize}
   \begin{center}
    \includegraphics[width=7cm,height=7cm,keepaspectratio]
{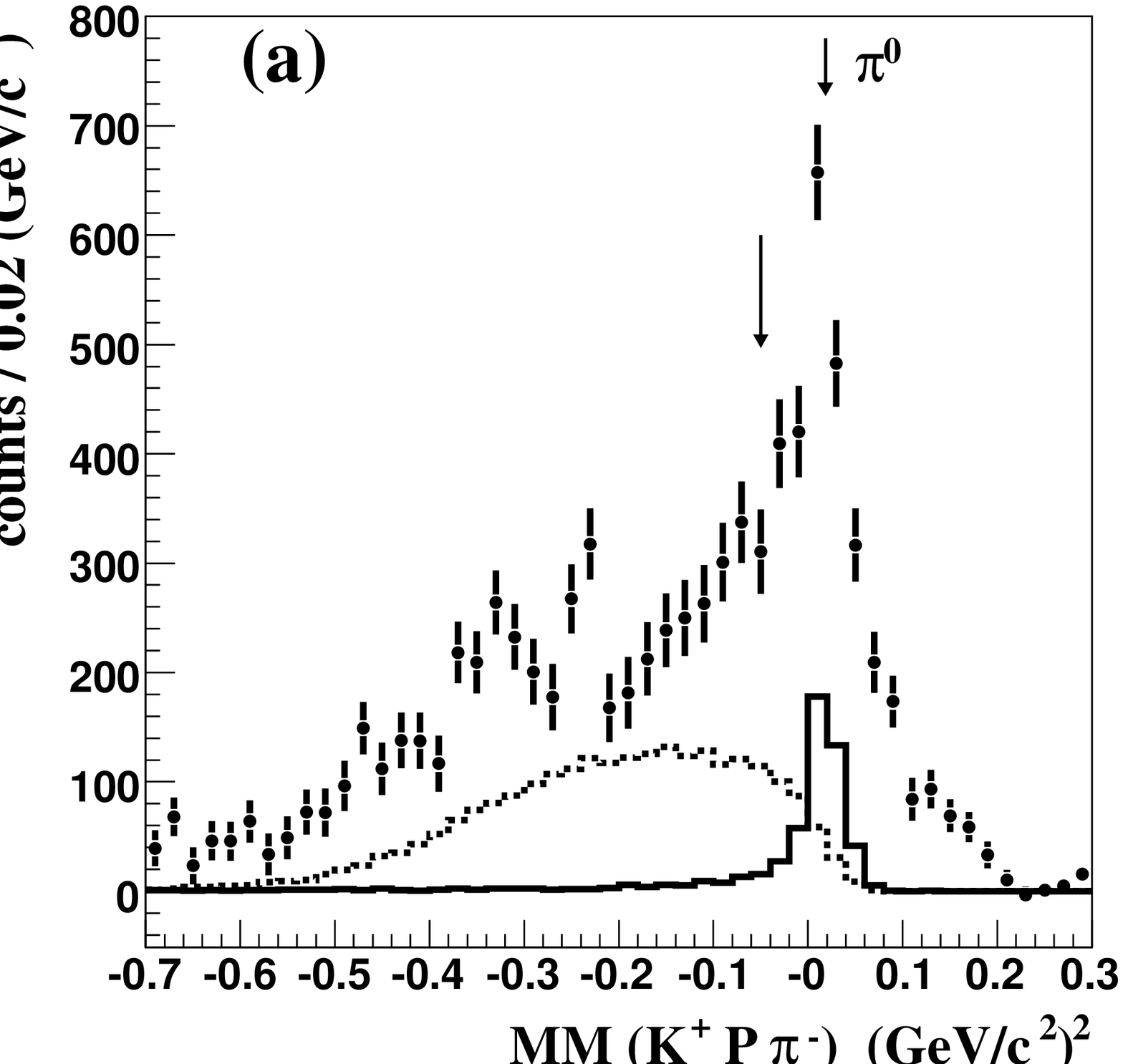}
   \end{center}
  \end{minipage} & %
  \begin{minipage}{0.5\hsize}
   \begin{center}
    \includegraphics[width=7cm,height=7cm,keepaspectratio]
{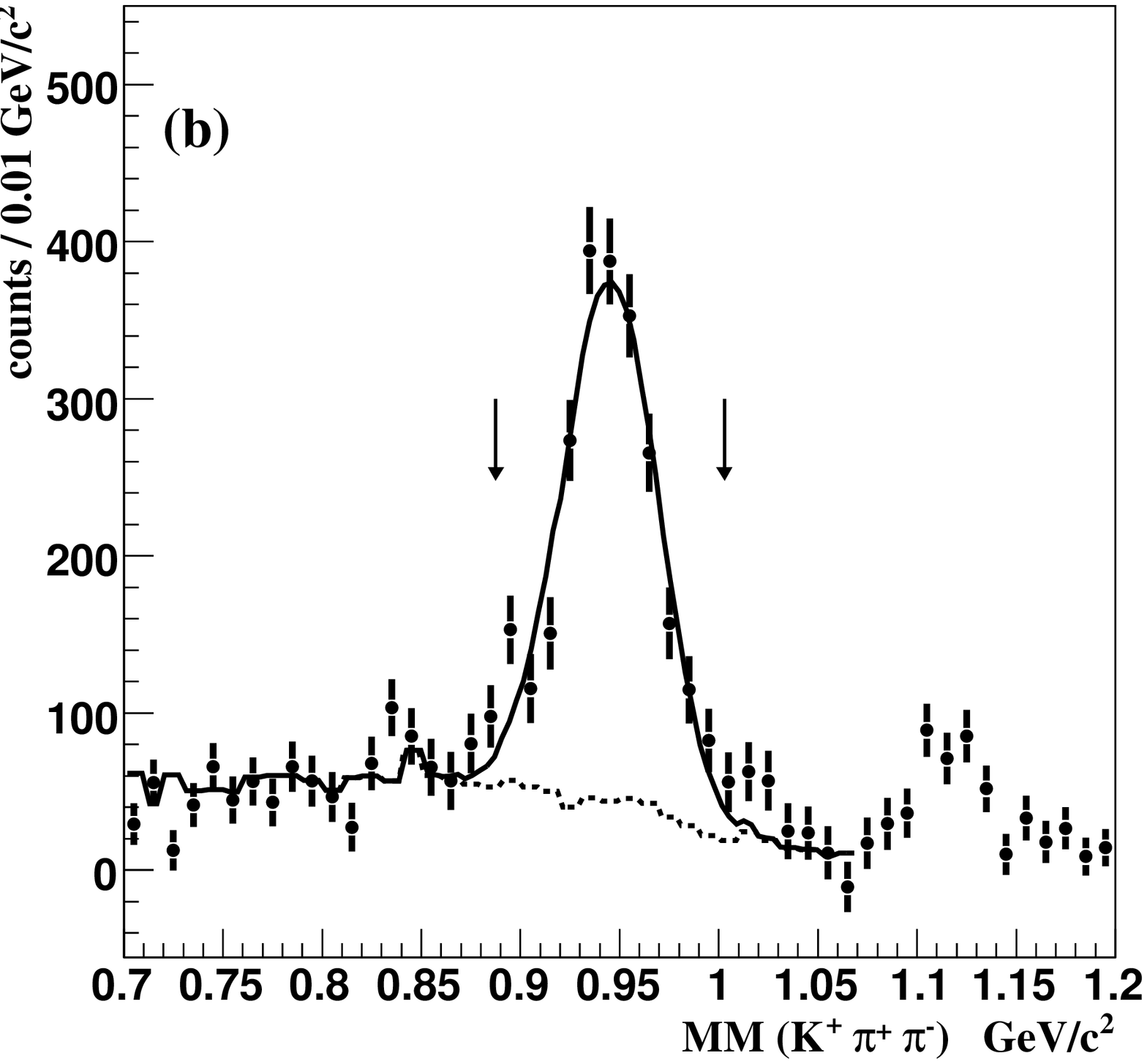}
   \end{center}
  \end{minipage} \\
  \begin{minipage}{0.5\hsize}
   \begin{center}
    \includegraphics[width=7cm,height=7cm,keepaspectratio]
{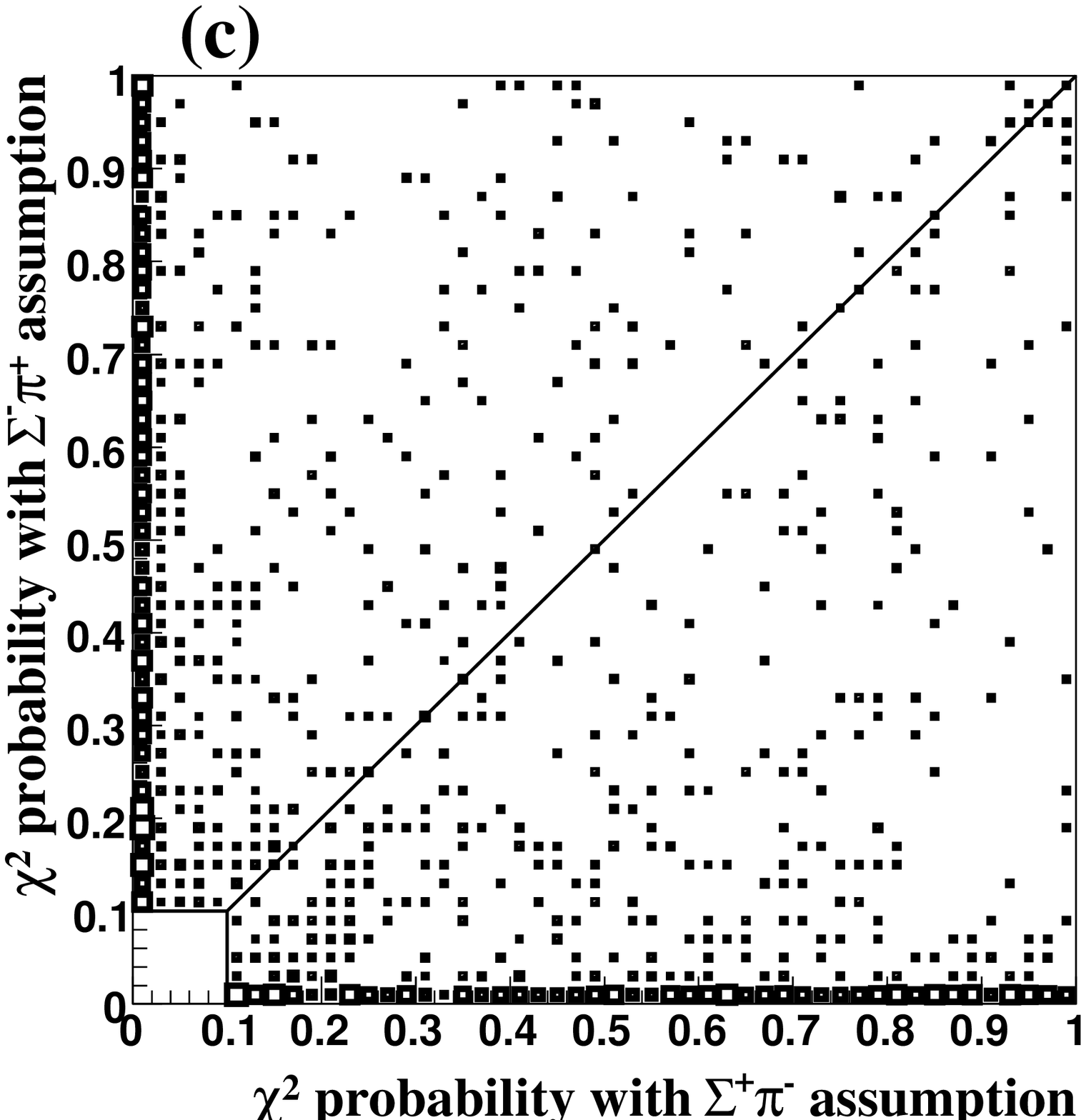}
   \end{center}
  \end{minipage} & %
  \begin{minipage}{0.5\hsize}
   \begin{center}
    \includegraphics[width=7cm,height=7cm,keepaspectratio]
{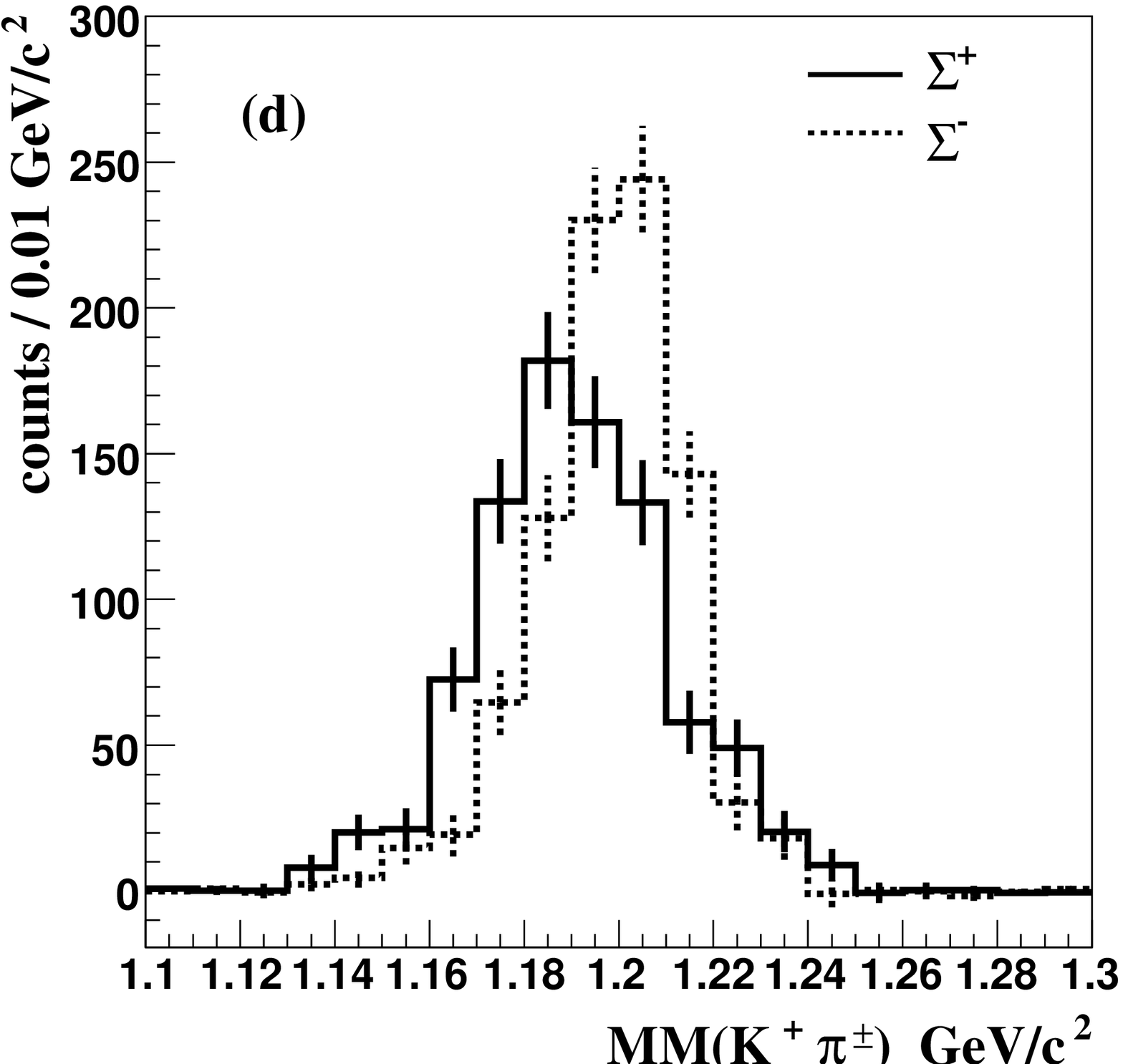}
   \end{center}
  \end{minipage} \\
 \end{tabular}
  \caption{(a) $MM^2( K^+ p \, \pi^-)$ distribution with an assumption
  that all $\pi^+$ candidates are protons.
The solid histogram shows the expected spectrum
 for the background reaction of 
 $\Sigma^0(1385)\rightarrow \Lambda \pi^0\rightarrow p \pi^- \pi^0$
 as generated by the MC simulation.
 The expected spectrum from the signal of \LamS production 
 is displayed as the dashed histogram.
  (b) $MM(K^+ \pi^+ \, \pi^-)$ for the $\gamma p \rightarrow K^+ \pi^+\pi^-X$ reaction. 
  (c) Correlation plot of the $\chi^2$ probability with
  $\Sigma^-\pi^+$ assumption and with $\Sigma^+\pi^-$ assumption.
  The solid line shows the boundary for the $\Sigma^+$/$\Sigma^-$ selection cuts.
  (d) $MM( K^+ \pi^-)$ (solid) and 
  $MM(K^+ \pi^+)$ (dashed) after the $\Sigma^+$ selection cut
  and the $\Sigma^-$ selection cut, respectively.
  } 
  \label{chi2pr_sigma}
\end{figure}

 The candidate tracks of a ($\pi^+$, $\pi^-$) pair were again identified by the
 energy deposition in the TPC combined with the momentum information.
 The $\pi^+$ selection boundaries were the same as those for the
 $\pi^-$'s as indicated by the solid lines in Fig.~\ref{PID_Lambda} (a).
 The rate at which protons are misidentified as $\pi^+$'s increases for
 the high momentum region, because $dE/dx$ for these protons becomes small
 and the proton band cannot be separated from the pion band.
 These protons were produced by either the
 $\gamma \, p \rightarrow K^+ \Sigma^0(1385)\rightarrow K^+ \Lambda \pi^0
 \rightarrow p \pi^- \pi^0$ reaction or 
$\gamma \, p \rightarrow K^+ \Lambda(1405)\rightarrow K^+ \Sigma^+
 \pi^-\rightarrow K^+ p \pi^0 \pi^-$ reaction.
 In both reactions, the missing mass of the $\gamma p\rightarrow K^+ p \pi^- X$
 reaction, $MM(K^+p \pi^-)$, corresponds to the $\pi^0$ mass.
 Thus,  these background events can be eliminated by rejecting $\pi^0$'s
 in $MM(K^+p \pi^-)$.
 Fig.~\ref{chi2pr_sigma} (a) shows the distribution of the square of
 $MM (K^+  p \pi^-)$, where all $\pi^+$ candidates were assumed to be
 protons and assigned the proton mass.
 Closed circles show the experimental data. 
 The arrows indicate the PDG value of the $\pi^0$ mass squared and the cut
 point for $\pi^0$ rejection. The enhancement
 of the spectrum at the $\pi^0$ mass due to the background reactions is
 clearly visible. The solid histogram shows the expected spectrum
 for the background reaction of 
 $\Sigma^0(1385)\rightarrow \Lambda \pi^0\rightarrow p \pi^- \pi^0$
 as generated by the MC simulation.
 The contamination from this reaction was reduced by requiring
 $ MM^2(K^+ \, p \, \pi^-) < -0.05$~(GeV/c$^2$)$^2$.
 The expected spectrum from the signal of \LamS production 
 is displayed as the dashed histogram.
 The acceptance of the cut $ MM^2(K^+ \, p \, \pi^-) < -0.05$ for
 \LamS was estimated to be 85\% by the MC.

 Neutrons in $\Lambda(1405)\rightarrow \Sigma^\pm\pi^\mp\rightarrow \pi^+\pi^-n$
 decay were identified from the missing mass of the
 $\gamma p \rightarrow K^+ \pi^+ \pi^- X$ reaction.
 Fig.~\ref{chi2pr_sigma} (b) shows $MM(K^+ \pi^+ \pi^-)$ for events
 which survived cut conditions (i) to (iii).
 A peak corresponding to the neutron mass is observed over a
 broad background.
 This background consists of events where a proton or a $\pi^+$
 was misidentified as a $K^+$ by the LEPS spectrometer, and the
 distribution of these background events is shown as the
 dashed line in Fig.~\ref{chi2pr_sigma} (b). 
 The contamination
 of this background into the neutron mass region was estimated to be 8\%.
 The mass peak position and width obtained from a Gaussian fit
 were $945 \pm 1$~MeV/c$^2$ and
 $19 \pm 2$~MeV/c$^2$, respectively.
 The width was consistent with the expected value of 20 MeV/c$^2$
 but the mass peak was shifted slightly.
 Events within $\pm 3 \sigma$ of the mass peak were kept for
 further analysis.

 A kinematic fit with two constraints (C2-fit),  
 $MM (K^+ \pi^+ \pi^-) = M(n)$
 and  $MM (K^+ \pi^{\pm}) = M(\Sigma^{\mp})$,
 was applied to purify the \LamS production events and
 to separate its $\Sigma^+\pi^-$ and $\Sigma^-\pi^+$ decay modes.
 The kinematic fit was applied twice for
 each event, once with the $\Sigma^+\pi^-$ decay assumption and
 again with the $\Sigma^-\pi^+$ decay assumption.
 Fig.~\ref{chi2pr_sigma} (c) shows the correlation of the
 $\chi^2$ probability of C2-fit between the $\Sigma^-\pi^+$
 assumption ($prob(\chi^2)_{\Sigma^-}$)
 and $\Sigma^+\pi^-$ assumption ($prob(\chi^2)_{\Sigma^+}$)
 for events which survived the event selection criteria of (i) to (iv) and
 the cut ($prob(\chi^2)_{\Sigma^+}>0.1$ or
 $prob(\chi^2)_{\Sigma^-}>0.1$).
 Events with a ($\Sigma^+\pi^-$) pair should have larger values for the
 $\chi^2$ probability with the $\Sigma^+\pi^-$ decay assumption than
 with the $\Sigma^-\pi^+$ decay assumption.
 As shown in Fig.~\ref{chi2pr_sigma} (c), events with large
 $prob(\chi^2)_{\Sigma^+}$ and $prob(\chi^2)_{\Sigma^-}$ are
 exclusive.
 In this way, we could distinguish the two decay modes using the
 $\chi^2$ probability of the C2-fit.
 The $\Sigma^+\pi^-$ decay events were selected
 by requiring $prob(\chi^2)_{\Sigma^+}>0.1$ and
 $prob(\chi^2)_{\Sigma^+}>prob(\chi^2)_{\Sigma^-}$, whereas the
 $\Sigma^-\pi^+$ decay events were selected by requiring
 $prob(\chi^2)_{\Sigma^-}>0.1$  and
 $prob(\chi^2)_{\Sigma^-}>prob(\chi^2)_{\Sigma^+}$.
 The solid line in  Fig.~\ref{chi2pr_sigma} (c) shows the boundary
 of the $\Sigma^+$/$\Sigma^-$ selection cut.
 The misidentification rate of $\Sigma^+$ and $\Sigma^-$ using the
 above procedure was estimated to be 12\% using MC simulation.
 The distributions of $ MM (K^+ \pi^{\pm})$ are shown in
 Fig.~\ref{chi2pr_sigma} (d). The solid histogram is 
 $MM (K^+ \pi^{-})$ and the dashed one is $MM(K^+ \pi^{+})$.
 The masses of $\Sigma^+(1189)$ and $\Sigma^-(1197)$ were
 determined via a Gaussian fit to the data to be
 $1191 \pm 1$ MeV/c$^2$ and $1199 \pm 1$ MeV/c$^2$, respectively.
 The measured widths of $\Sigma^+(1189)$ and $\Sigma^-(1197)$ were 
 $20 \pm 1$ MeV/c$^2$ and $16 \pm 1$ MeV/c$^2$, and are
 consistent with the expected value of 17 MeV/c$^2$ as estimated by MC.

 The measured spectra for the $\Sigma^+\pi^-$ and $\Sigma^-\pi^+$ modes
 were compared with each other and with spectra from a previous
 measurement~\cite{Ahn2003}.
 In the previous measurement, both a $K^+$ and a charged pion
 were detected in the LEPS spectrometer. On the other hand, in this work, 
 a $K^+$ was detected in the LEPS spectrometer, and two charged pions
 were measured by the TPC. Therefore, these two measurements differ in
 the angle between the $K^+$ and the pion.
 Fig.~\ref{L1405} (a) and (b) show the spectrum of $MM(K^+)$ after the
 $\Sigma^+$ and $\Sigma^-$ selection cuts, respectively.
 The spectra obtained by this work are shown as closed circles. 
 Open circles show the unnormalized spectra from the previous measurement \cite{Ahn2003}.
 The $\Lambda(1520)$ peak visible in these spectra was fitted 
 using a Breit-Wigner function atop the phase space distribution of
 nonresonant ($K^+\Sigma\pi$) production.
 The solid lines show the fit results.
 The mass peak positions are $1520 \pm 2$ MeV/c$^2$
 in the $\Sigma^+\pi^-$ decay mode and $1517 \pm 2$ MeV/c$^2$
 in the $\Sigma^-\pi^+$
 decay mode. Thus, the mass of $\Lambda(1520)$ is consistent
 with the PDG value in each decay mode.
 The peak position of the \LamS in $\Sigma^-\pi^+$ was consistent
 with the PDG value of 1405 MeV/c$^2$.
 However, the peak structure in the $\Sigma^+\pi^-$ mode was not clear.
 The decay mode dependence of the lineshapes of \LamS is likely due to
 strong interference between isospin 0 and 1 amplitudes of the $\Sigma\pi$
 interaction, as discussed in Ref.~\cite{Nacher1999}. 
 The apparent difference for the lineshape of the \LamS in the $\Sigma^-\pi^+$
 decay mode between the current work and the previous measurement will be 
 discussed in the next section.
 The isospin interference term is canceled by summing the spectra of
 the $\Sigma^+\pi^-$ and $\Sigma^-\pi^+$ modes.
 The summed spectrum was obtained after correcting for the decay branch of
 $\Sigma^+\rightarrow p\, \pi^0$ ($\sim 52$\%), and the result is shown in Fig.~\ref{L1405} (c). 
 Closed and open circles show the spectra measured by this work and 
 by the previous one, respectively, where the normalization for the spectrum
 by the previous measurement was determined by fitting
 in the range of $1.34<MM(K^+)<1.47$ GeV/c$^2$. The $\chi^2$/ndf was 1.4.
 Thus, the lineshape of \LamS after the sum is consistent with the one from the
 previous measurement.

 The yield of \LamS was extracted by fitting the theoretical spectrum of 
 Nacher {\it et al.}~\cite{Nacher1999} to the peak in the combined
 spectrum of the $\Sigma^+\pi^-$ and $\Sigma^-\pi^+$ modes.
 The combined spectrum is shown as closed circles in Fig.~\ref{L1405-2} for
 $0.8<\cos(\Theta_{K_{CM}})<1.0$ and two photon energy ranges:
 $1.5<E_\gamma<2.0$~GeV (a) and $2.0<E_\gamma<2.4$~GeV (b).
 The spectra were corrected for the detector acceptance and were normalized
 using the differential cross section of
 $K^+\Lambda(1116)$ production measured from data set (I)~\cite{Sumihama2006}
 in each photon energy bin.
 The spectra were fitted with the distribution for $K^+\Lambda(1405)$, 
 $K^+\Lambda(1520)$ and nonresonant ($K^+\Sigma\pi$) production as determined
 by MC simulation.
 The strength of each reaction was obtained by the fitting, with the assumption
 that the ratio of the yields of nonresonant ($K^+\Sigma\pi$) production in the
 two photon energy regions is proportional to the phase volume.
 The solid curves show the spectra of \LamS calculated by Nacher {\it et al.},
 and the dashed lines show the distribution for nonresonant ($K^+\Sigma\pi$)
 production.
 The contamination from ($K^{*0}\Sigma^+$) production was
 measured using the invariant mass distribution of ($K^+\pi^-$) pairs in the
 $2.0<E_\gamma<2.4$~GeV region, and 
 the expected spectrum of ($K^{*0}\Sigma^+$) production generated by the
 MC simulation is shown as the dot-dashed line in Fig.~\ref{L1405-2} (b).
 The open circles show the spectrum of $K^+\Sigma^{0}(1385)$ production with
 normalization determined from the yield found above.
 The fit results are shown as the solid histograms. 
 The $\chi^2/ndf$ for the fits were 1.8 and 1.7 for photon energy
 of $1.5 < E_\gamma < 2.0 $ GeV and  $2.0 < E_\gamma < 2.4 $ GeV, respectively.
 The theoretical spectrum of Nacher {\it et al.} is seen to be consistent
 with the experimental data in the low photon energy region.
 A second fit was performed using a different theoretical spectrum
 due to Kaiser~{\it et al.}~\cite{Kaiser1995} derived from an effective
 Lagrangian.
 The fit results didn't change significantly, and 
 this theoretical model is also seen to be consistent with
 the experimental data.
 In the high photon energy region, the lineshape of \LamS is unclear, and 
 the yield of \LamS extracted by fitting depends on the estimation of the background
 reaction. A more conservative yield estimation will be discussed
 later.

 After correcting for the detector acceptance and decay branches of
 the hyperon resonances, the production ratios of \LamS to \SigS were obtained
 as $ \Lambda^*/\Sigma^* =  0.54 \pm 0.17 $ and $0.084 \pm 0.076 $
 for  $1.5 < E_\gamma < 2.0 $ GeV and $2.0 < E_\gamma < 2.4$ GeV, respectively.
 The systematic uncertainties due to the detection efficiency of the TPC,
 the target thickness and the number of photons were canceled out
 in the ratio.


 Finally, the absolute values of the differential cross sections of \LamS 
 and \SigS production off protons were measured from data set (I) using
 the production ratio of these two hyperons determined above.
 The event selection criteria were the same as for the analysis of the
 $\gamma p \rightarrow K^+ \Lambda/\Sigma^0$ production reactions. 
 Details can be found in Ref.~\cite{Sumihama2006}.
 The angular coverage for forward going $K^+$'s was matched with
 that of data set (II) by selecting the overlapping region,
 $0.8<\cos\Theta_{K_{CM}}<1.0$.

\begin{figure}[htb]
\begin{tabular}{c c}
   \begin{minipage}{0.5\hsize}
    \begin{center}
     \includegraphics[width=7.5cm,height=7.5cm,keepaspectratio]
     {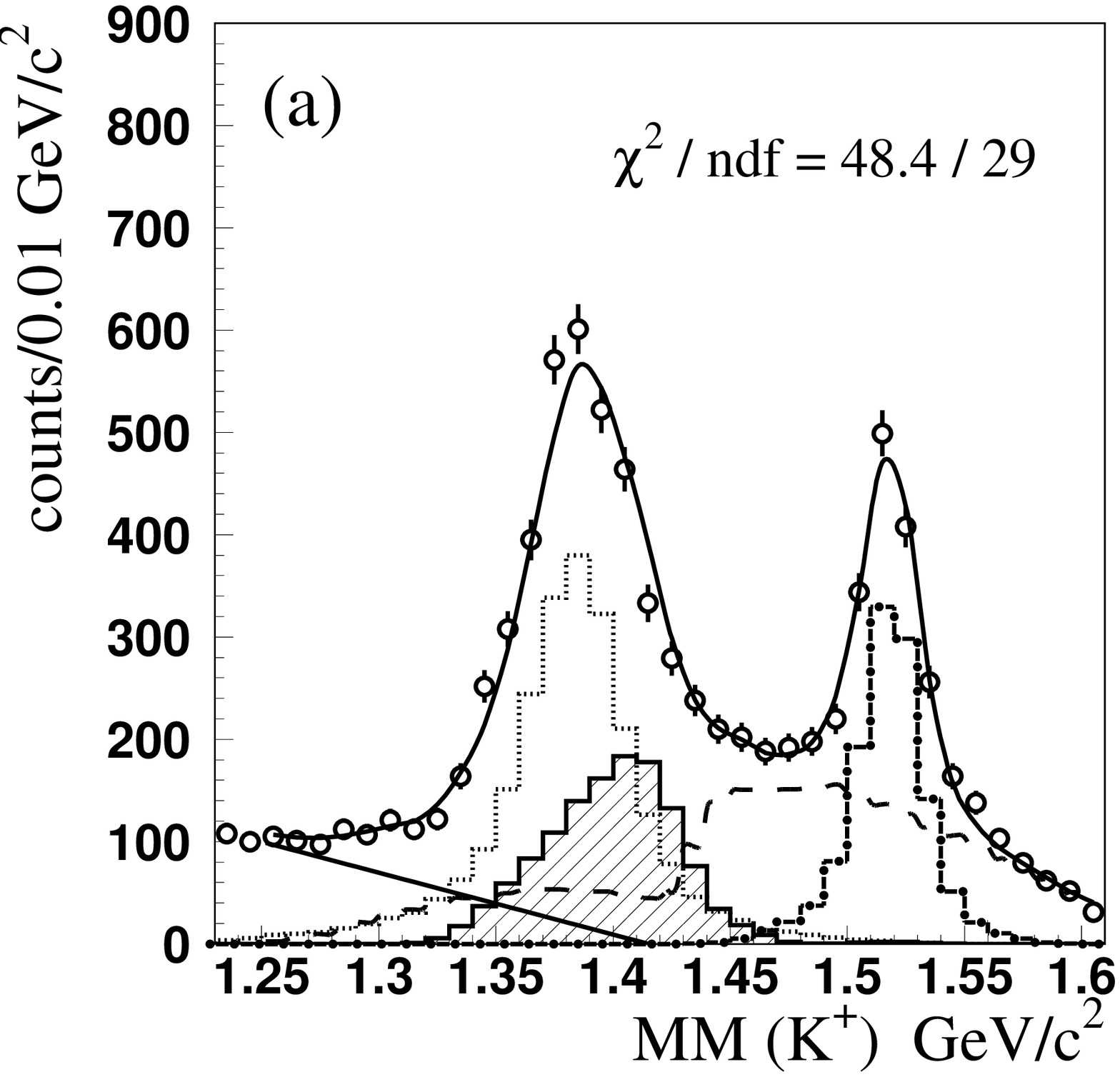}
    \end{center}
   \end{minipage} & %
   \begin{minipage}{0.5\hsize}
    \begin{center}
     \includegraphics[width=7.5cm,height=7.5cm,keepaspectratio]
     {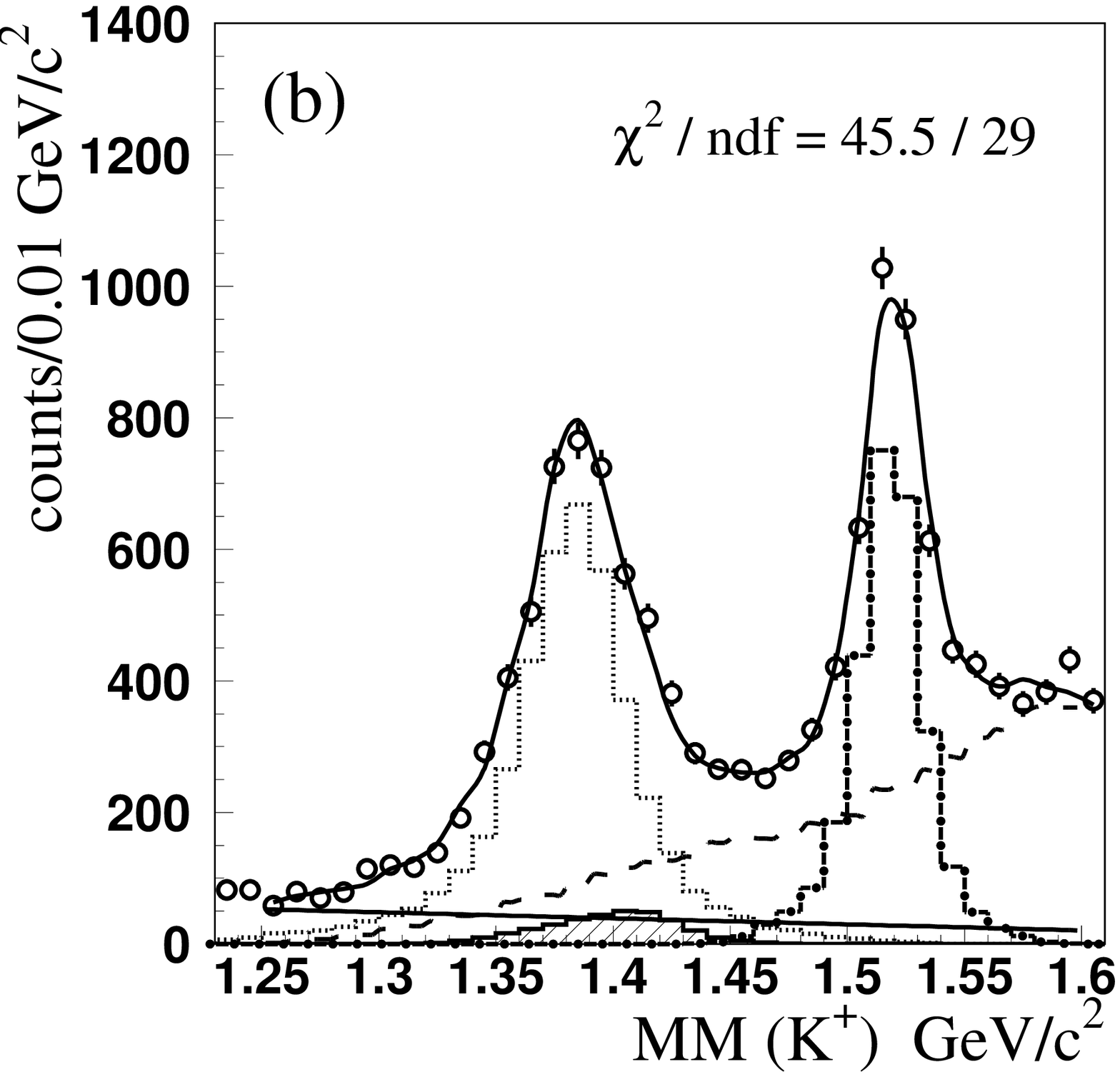}
    \end{center}
   \end{minipage} \\
\end{tabular}
   \caption{Missing mass spectra of the $\gamma p \rightarrow K^+ X$ reaction for
   photon energy ranges of (a) $1.5 < E_\gamma < 2.0$ GeV and
   (b) $2.0 < E_\gamma < 2.4$ GeV in data set (I).
   The data from hydrogen target is shown as open circles.
   The solid histograms show fit results. 
   The hatched, dotted and dot-dashed histograms are the spectra of \LamS, \SigS
   and $\Lambda(1520)$, respectively.
   The dashed lines are summed spectra of background reactions of 
   ($K^+\Lambda \, \pi$), ($K^+\Sigma \, \pi$), ($K^+K^-p$) and
   $\phi$-meson production.
   The additional background contributions which could not be reproduced
   by the sum of the known physics processes are fitted with a linear function
   and indicated by the straight lines.
   }
   \label{L1405_slh2}
  \end{figure}

 Fig.~\ref{L1405_slh2} shows $MM(K^+)$ from the liquid hydrogen target
 for $0.8<\cos\Theta_{K_{CM}}<1.0$ and the two photon energy ranges,  
 $1.5 < E_\gamma < 2.0$ GeV (a) and $2.0 < E_\gamma < 2.4$ GeV (b)
 in data set (I).
 The experimental data (open circles) were fitted with distributions
 for $\Lambda(1405)$ (hatched), $\Sigma^0(1385)$
 (dotted), $\Lambda(1520)$ (dot-dashed)
 and background reactions (dashed).
 The background reactions considered were
 nonresonant ($K^+\Lambda\pi$), ($K^+\Sigma\pi$),
 ($K^+K^-p$) and $\phi$-meson production.
 The normalization factor of each background spectrum
 was determined by fitting, and the sum of these background
 spectra are shown.
 The spectral shape of \LamS was assumed to be the one of the
 theoretical calculation by Nacher {\it et al.}~\cite{Nacher1999}. 
 A linear background was introduced to explain the background 
 events at the threshold of ($K^+\Lambda\pi^0$) production, 1.25 GeV/c$^2$, 
 where the contribution from $K^+\Sigma^0$ production was negligible.
 These background events might be caused by the mismeasurement of
 the photon energy or the $K^+$ momentum or near threshold
 enhancement of ($K^+\Lambda\pi^0$) production which
 was not included in the MC simulation.
 The main systematic uncertainties due to this background were estimated
 to be $^{+1.0}_{-27}$\% and $^{+8.1}_{-0.94}$\%
 for $1.5<E_\gamma<2.0$ GeV and $2.0<E_\gamma<2.4$ GeV, respectively,
 by fitting with various slope parameters of the linear background
 and without the linear background.
 The other sources of systematic uncertainties are summarized in Table~\ref{sysErr}.
 The differential cross sections of \LamS production were found to be
 $d\sigma/d(\cos\theta) = 0.43 \pm 0.088(stat.)^{+0.034}_{-0.14} (syst.)$ 
 $\mu$b and $ 0.072 \pm 0.061(stat.)^{+0.011}_{-0.0056}(syst.)$ $\mu$b for
 $1.5<E_\gamma<2.0$ GeV and  $2.0<E_\gamma<2.4$ GeV, respectively.
 Those of \SigS production were
 $ 0.80 \pm 0.092(stat.)^{+0.062}_{-0.27}(syst.)$ $\mu$b and
 $ 0.87 \pm 0.064(stat.)^{+0.13}_{-0.067}(syst.)$ $\mu$b for
 $1.5<E_\gamma<2.0$ GeV and $2.0<E_\gamma<2.4$ GeV, respectively.

  {\small
  \begin{table}[tbh]
   \begin{center}
    \caption{The sources of systematic uncertainties for the measurement of differential
    cross sections.}
    \label{sysErr}
\begin{tabular}{l r r}
\hline 
The sources of uncertainties & $1.5<E_{\gamma}<2.0$ GeV & $2.0<E_{\gamma}<2.4$ GeV \\
Background around ($K^+\Lambda\pi$) threshold & $^{+1.0}_{-27}$ \% & $^{+8.1}_{-0.94}$ \% \\ 
The sources of uncertainties  & \multicolumn{2}{c}{$1.5<E_{\gamma}<2.4$ GeV} \\
Thickness of the liquid H$_2$ target    & \multicolumn{2}{r}{1.0 \%}\\ 
Number of photons  & \multicolumn{2}{r}{1.2 \%} \\ 
Photon transmission efficiency & \multicolumn{2}{r}{3.0 \%}\\ 
Accidental veto by   &   & \\ 
\qquad the aerogel \v{C}erenkov counter & \multicolumn{2}{r}{1.6 \%}  \\ 
\hline 
\end{tabular} 

   \end{center}
  \end{table}
  }
  
\section{Discussion}

\begin{figure}[htbp]
\begin{tabular}{c c}
  \begin{minipage}{0.5\hsize}
   \begin{center}
    \includegraphics[width=6.0cm,height=6.0cm,keepaspectratio]
{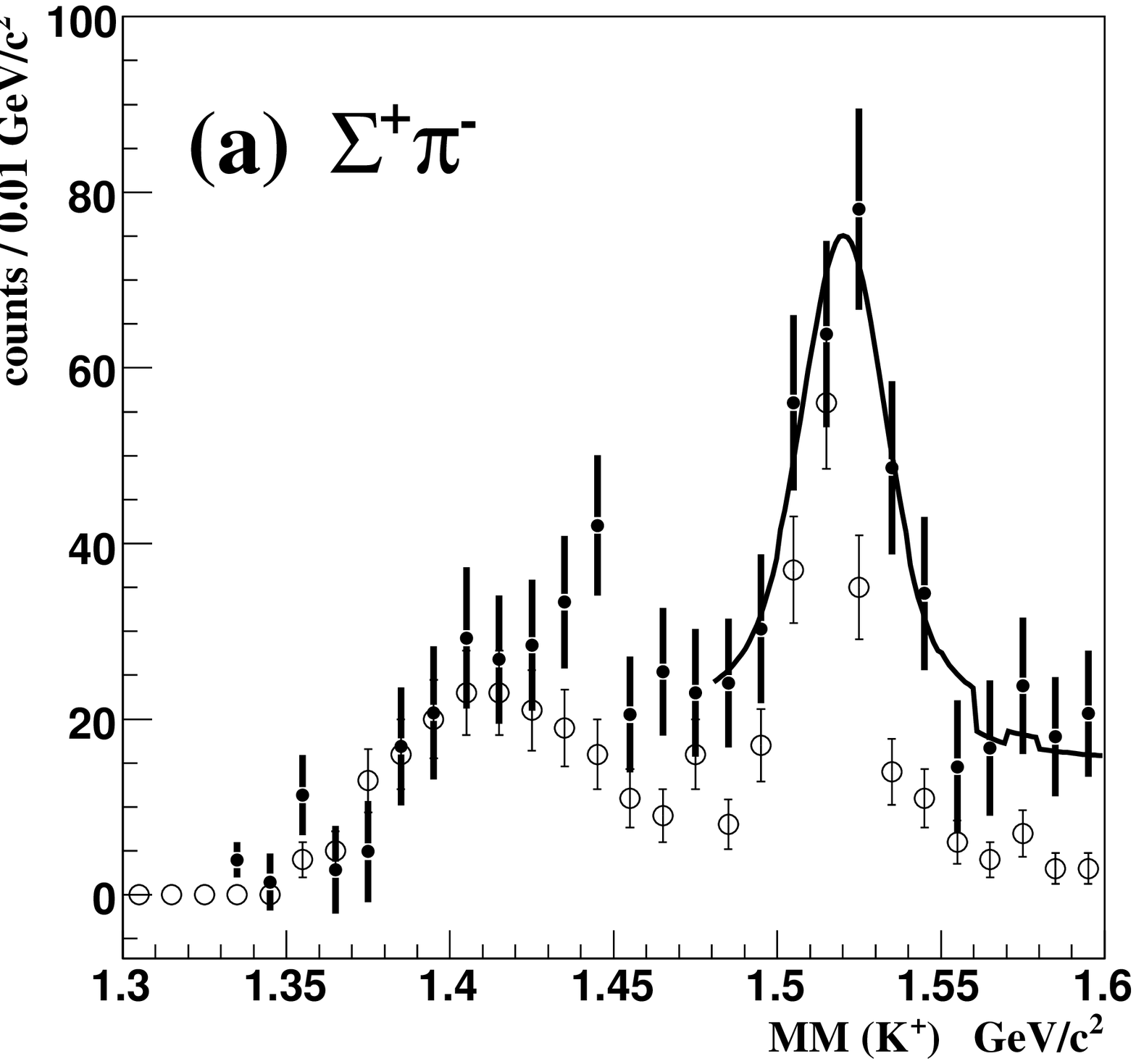}
   \end{center}
  \end{minipage} & %
  \begin{minipage}{0.5\hsize}
   \begin{center}
    \includegraphics[width=6.0cm,height=6.0cm,keepaspectratio]
{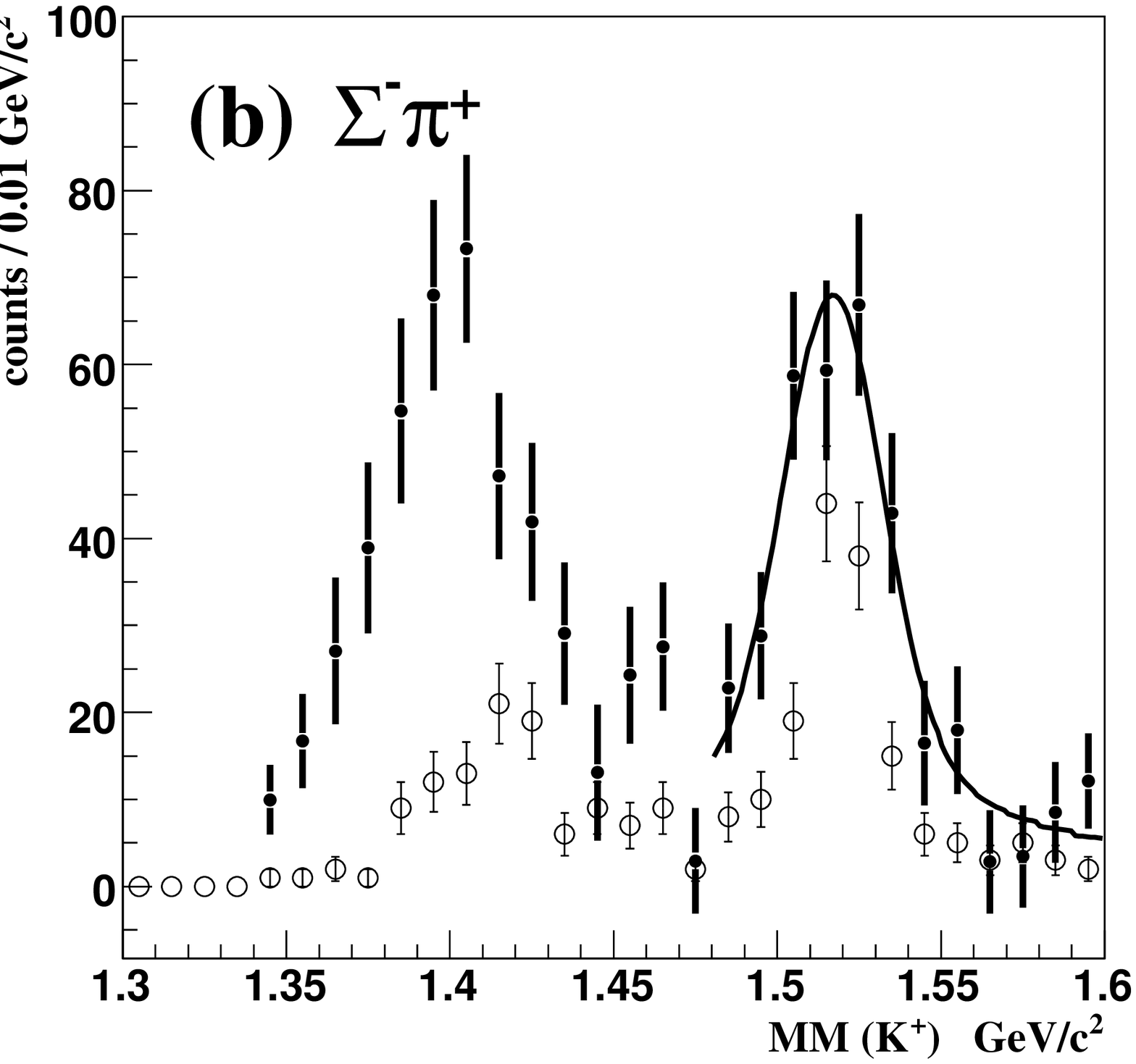}
   \end{center}
  \end{minipage} \\
  \begin{minipage}{0.5\hsize}
   \begin{center}
    \includegraphics[width=6.0cm,height=6.0cm,keepaspectratio]
{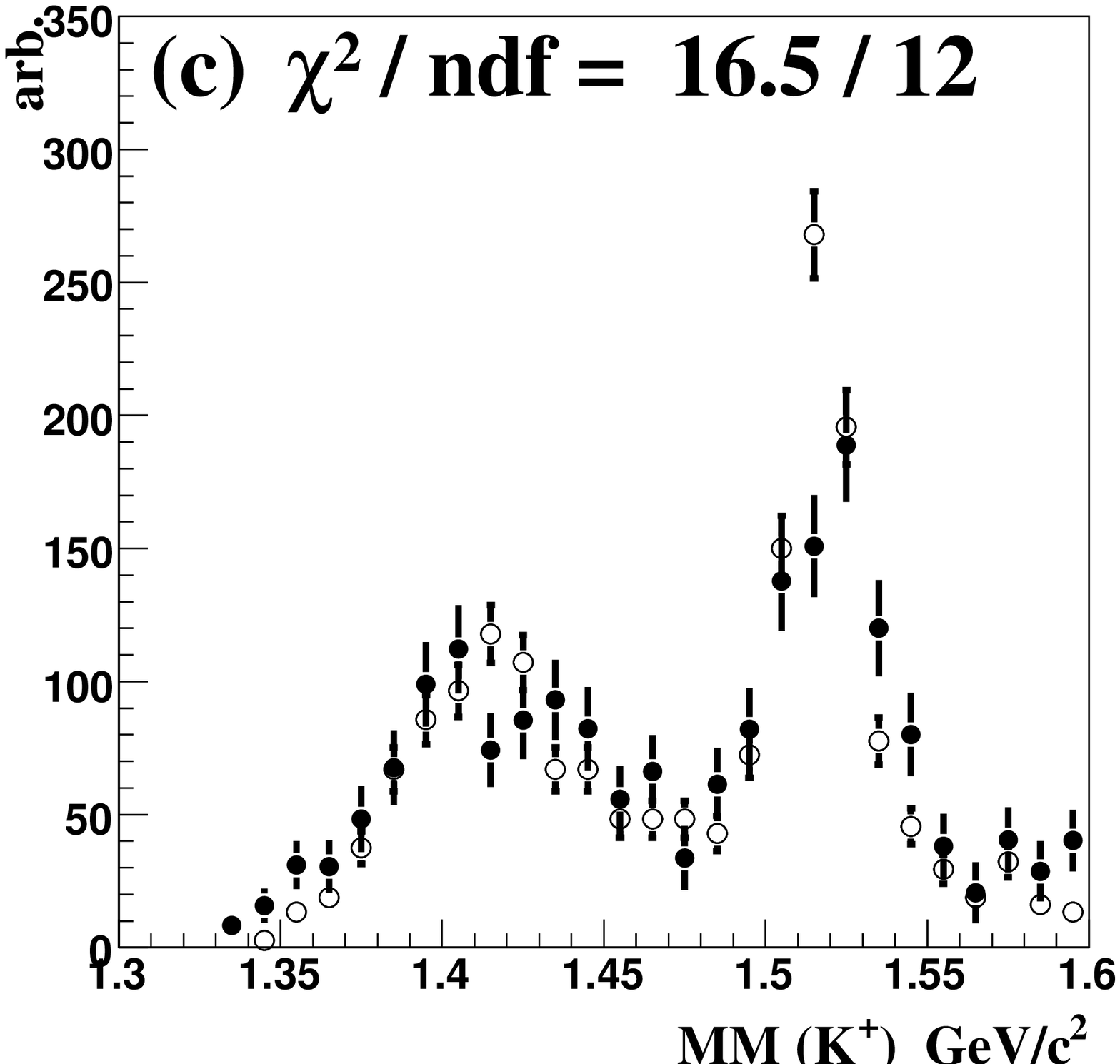}
   \end{center}
  \end{minipage} & \\
\end{tabular}
  \caption{Missing mass for the $\gamma p \rightarrow K^+X$ reaction.
  (a) $K^+\Sigma^+ \pi^-$ final state.
  (b) $K^+\Sigma^- \pi^+$ final state.
  Solid lines in (a) and (b) show fit results
  of $K^+\Lambda(1520)$ plus nonresonant ($K^+\Sigma\pi$) production.
  (c) The combined spectra of the $\Sigma^+ \pi^-$ and $\Sigma^- \pi^+$ decay modes.
  Closed and open circles show spectra obtained by this work and by a previous
  measurement~\cite{Ahn2003}, respectively. 
  }
  \label{L1405}
 \end{figure}

 The observed mass spectra of \LamS in $\Sigma^+\pi^-$ mode and in $\Sigma^-\pi^+$
 mode shown in Fig.~\ref{L1405} are distinct, which is understood to be due to
 the interference between the isospin 0 and 1 amplitudes~\cite{Nacher1999}.
 In addition, the lineshapes of \LamS in the $\Sigma^-\pi^+$ decay mode
 measured by this work and by the previous work are different with each other
 as shown by the closed and open circles in Fig.~\ref{L1405} (b).
 The photon energy and the scattering angles of the $K^+$ were the same in
 these two measurements. However, the range of polar angles for the pion with respect to
 the momentum vector of the $\Lambda(1405)$ was quite different. 
 In the previous measurement, both a $K^+$ and a charged pion were detected in
 the LEPS spectrometer. The direction of the \LamS is opposite to that
 of the $K^+$ in the center-of-mass frame, and thus, the direction of
 the pion relative to the $\Lambda(1405)$, $\Theta_{\pi\Lambda^*}$,
 is near 180 degrees.
 According to MC simulation,  90\% of events fall in the range of
  $\cos(\Theta_{\pi\Lambda^*})<-0.4$.
 In this work, a $K^+$ was detected in the LEPS
 spectrometer and charged pions were measured in the TPC, which covers the
 side of the target and can access the entire range of $\Theta_{\pi\Lambda^*}$.
 Thus, the angles of the momentum vector of the pion
 relative to the direction of the \LamS in the center-of-mass frame
 were distributed over a much wider range than in the previous measurement.
 Therefore, the difference of the lineshapes between the two
 measurements might be explained by an angular dependence of the interference
 term of the $\Sigma\,\pi$ scattering amplitudes.

\begin{figure}[htbp]

\begin{tabular}{c c}
  \begin{minipage}{0.5\hsize}
   \includegraphics[width=6.0cm,height=6.0cm,keepaspectratio]
   {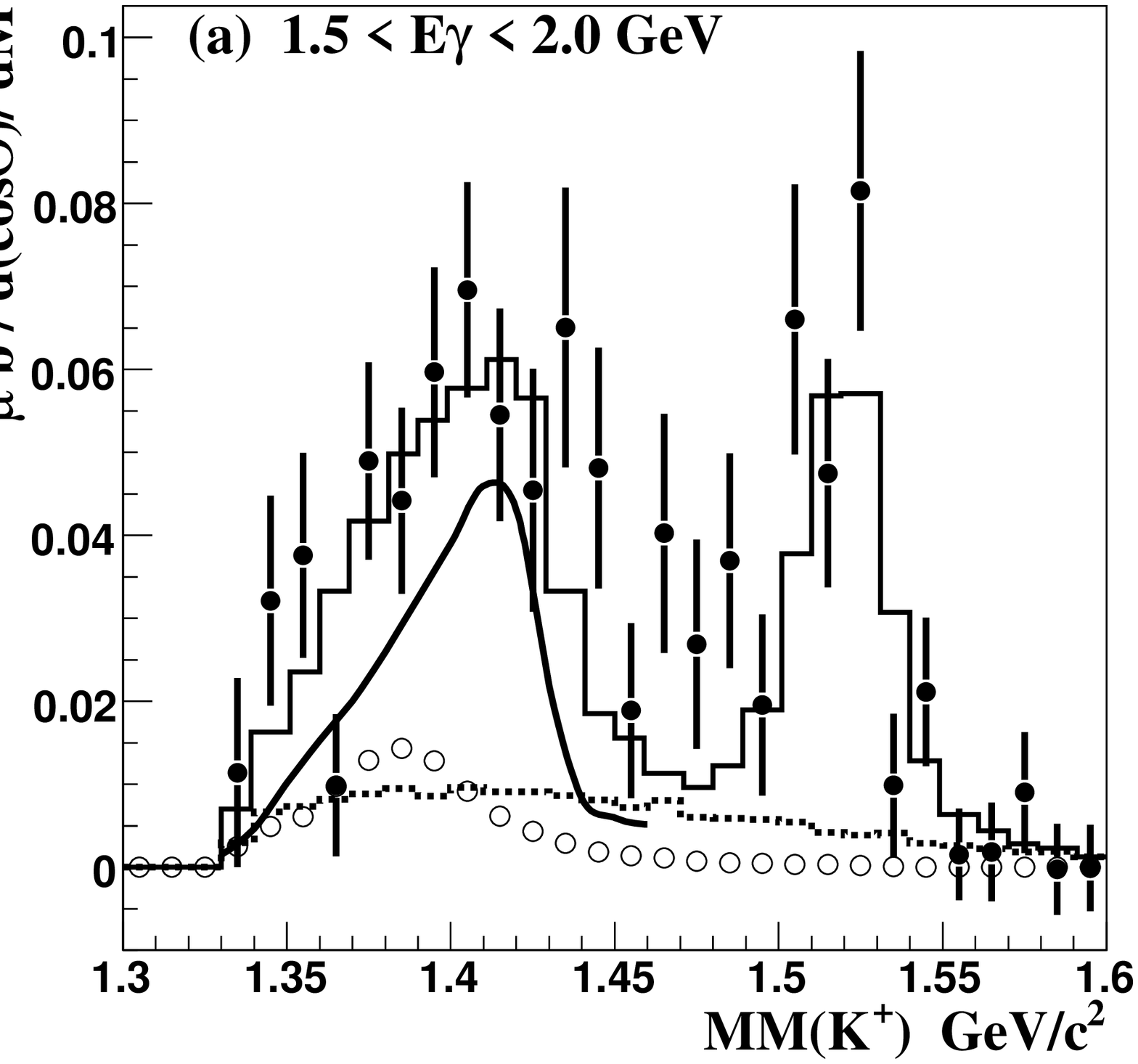}
  \end{minipage} & %
  \begin{minipage}{0.5\hsize}
   \includegraphics[width=6.0cm,height=6.0cm,keepaspectratio]
   {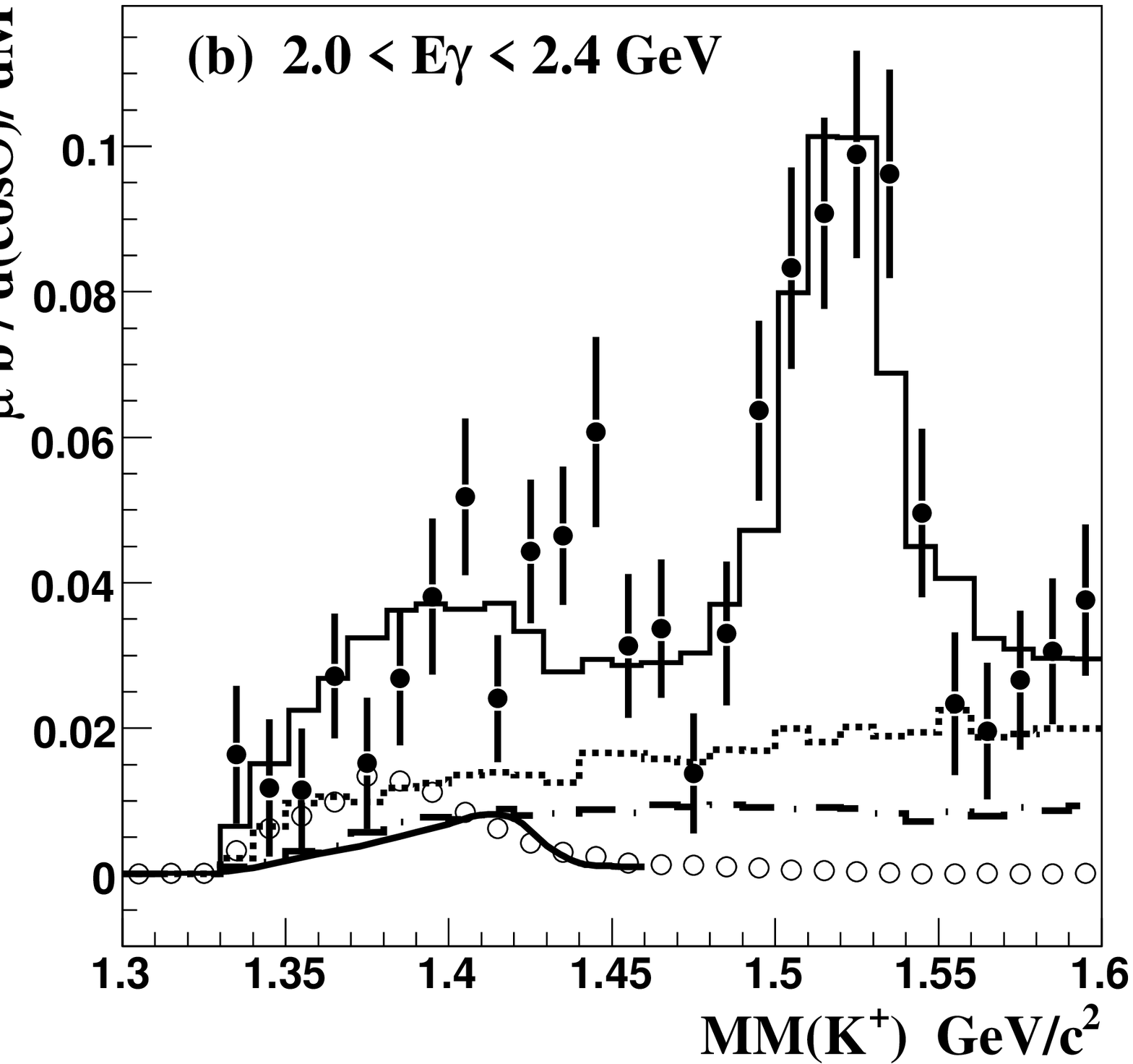}
  \end{minipage} \\
\end{tabular}
  \caption{ Missing mass of the $\gamma p \rightarrow K^+X$ reaction
  in data set (II) in two photon energy ranges: (a) $1.5<E_\gamma<2.0$ GeV and 
  (b) $2.0<E_\gamma<2.4$ GeV.
  The experimental data are shown as closed circles.
  The data were fitted with spectra determined by MC simulation of
  $K^+\Lambda(1405)$, $K^+\Sigma^{0}(1385)$, $K^+\Lambda(1520)$,
  nonresonant ($K^+\Sigma\pi$) and $K^{*0}\Sigma^+$ production.
  The solid histograms show fit results. 
  The solid lines, open circles, dashed lines and dot-dashed line 
  show spectra of $K^+\Lambda(1405)$, $K^+\Sigma^{0}(1385)$,
  nonresonant ($K^+\Sigma\pi$) and $K^{*0}\Sigma^+$ production,
  respectively.
  }
  \label{L1405-2}
 \end{figure}

 The production ratios of \LamS to \SigS and the differential cross sections
 for the photoproduction of these hyperon resonances were obtained for photon
 energies in the region near production threshold,
 $1.5 < E_\gamma < 2.0 $ GeV, and above,
 $2.0 < E_\gamma < 2.4$ GeV, for $K^+$ scattering angles in the range
 $0.8<\cos\Theta_{K_{CM}}<1.0$.
 The production ratios of \LamS to \SigS were obtained
 as $ \Lambda^*/\Sigma^* =  0.54 \pm 0.17 $ and $0.084 \pm 0.076 $
 for  $1.5 < E_\gamma < 2.0 $ GeV and $2.0 < E_\gamma < 2.4$ GeV, respectively.
 The production of \LamS decreased in the higher photon energy region with
 respect to that of $\Sigma^{0}(1385)$.
 However, the spectrum of \LamS in the higher photon energy region was
 unclear (as shown in Fig.~\ref{L1405-2}(b)), and the yield of \LamS
 found by fitting
 depends on the estimation of the strength of the background.
 To obtain a more conservative estimate, the integral of the invariant mass
 distribution was considered.
 First, the strength in the range of $1.33<MM(K^+)<1.44$ GeV/c$^2$,
 including $\Lambda(1405)$, \SigS and all background reactions, was
 compared in two photon energy bins. 
 The ratios of this combined strength to \SigS production
 were $0.80\pm 0.23$ and $0.50\pm0.14$ 
 for $1.5 < E_\gamma < 2.0 $ GeV and $2.0 < E_\gamma < 2.4$ GeV, respectively.
 The production cross section for \SigS slightly increases in the higher
 photon energy region, and the contributions of the ($K^+\Sigma\pi$) 
 and ($K^{*0}\Sigma^+$) production also increase for higher photon energy
 as the phase volume increases. 
 Therefore, even in the conservative estimation, the cross section of \LamS 
 production should decrease remarkably.
 Next, in order to estimate the cross section of \LamS production, we 
 subtracted the contamination of \SigS and $K^{*0}$ production. Although the
 amplitude of \LamS production can interfere with these background amplitudes, 
  we subtracted the square of \SigS and $K^{*}$ amplitudes because a
 theoretical model calculating such an interference term is not available.
 However, we note that the contribution of \SigS is not negligible even 
 in the lower photon energy region, and a theoretical study of the
 interference with the p-wave amplitude is necessary.
 The ratios of the sum of \LamS and nonresonant $(K^+\Sigma\pi)$ production to
 \SigS production were $0.65 \pm 0.19$ and $0.35\pm 0.12$ for
 $1.5 < E_\gamma < 2.0 $ GeV and $2.0 < E_\gamma < 2.4$ GeV, respectively.
 Thus, the production of \LamS relative to \SigS is seen to decrease in 
 in the higher photon energy region even without any specific knowledge of
 the $(K^+\Sigma\pi)$ contribution.
 
 In order to gauge the effect of the form factor of $\Lambda(1405)$,
 the momentum transfer 
 was calculated in each photon energy region.
 The four momentum transfer, $t$, was $\sim$-0.22~(GeV)$^2$
 for photon energy of $1.5 < E_\gamma < 2.0 $ GeV and $\sim$-0.20~(GeV)$^2$ for
 $2.0 < E_\gamma < 2.4$ GeV.
 The momentum transfer in the two photon energy bins is similar, so our
 measurement is not sensitive to the $t$ dependence of the \LamS cross section.
 Thus, the reduction of \LamS production is likely related to some other
 details of the production mechanism.
 The observed differential cross sections of \SigS photoproduction
 of $\sim 0.8$ $\mu$b in the two photon energy bins are
 consistent with the effective Lagrangian calculation of
 Oh {\it et al.}~\cite{Oh2007} (0.6-1.1~$\mu$b).
 The differential cross section of \LamS production was obtained to be
  $\sim 0.4$ $\mu$b in $1.5 < E_\gamma < 2.0 $ GeV and
 $0.8<\cos\Theta_{K_{CM}}<1.0$.
 Nacher {\it et al.} predicted the \LamS production cross section as
 $\sim 0.8$ $\mu$b at $E_\gamma = 1.7$ GeV in $-1<\cos\Theta_{K_{CM}}<1$.
 Although the angular dependence of the cross section is unknown,
 the order of magnitude of the \LamS production cross section
 is consistent with the theoretical prediction.

 In summary, we have measured the photoproduction of the \SigS and \LamS hyperon
 resonances from the $\gamma p \rightarrow K^+ Y^*$ reaction
 in the photon energy of 1.5-2.4 GeV and
 in the polar angle range of $0.8<\cos\Theta_{K_{CM}}<1.0$.
 The production of \SigS was measured in the $(K^+\Lambda\pi^0)$ final state to
 which \LamS is prohibited to decay by isospin conservation. 
 The \LamS hyperon was measured in the $(K^+\Sigma^\pm\pi^\mp)$ final state,
 where the contamination from \SigS was estimated from the ($K^+\Lambda\pi^0$) final
 state.
 The lineshapes of the \LamS in $\Sigma^+\pi^-$ and $\Sigma^-\pi^+$ mode were
 different, which indicates  strong interference between isospin
 0 and 1 amplitudes.
 The combined spectrum shape of the $\Sigma^+\pi^-$ and $\Sigma^-\pi^+$
 modes, for which the interference term was canceled, was consistent with
 the theoretical calculation by Nacher {\it et al.}~\cite{Nacher1999} and 
 Kaiser~{\it et al.}~\cite{Kaiser1995}.

 The production ratios of \LamS to \SigS were determined to be $ 0.54 \pm 0.17 $
 and $0.084 \pm 0.076 $ for  $1.5 < E_\gamma < 2.0 $ GeV and
 $2.0 < E_\gamma < 2.4$ GeV, respectively.
 The ratio decreased in the higher photon energy region, which may suggest that
 the production mechanisms and form factors of \LamS photoproduction are
 largely different from those of $\Sigma^0(1385)$.

 The differential cross sections of \SigS and \LamS photoproduction
 were obtained in the two photon energy bins, $1.5<E_\gamma<2.0$ GeV and
 $2.0<E_\gamma<2.4$ GeV, and in the range of $K^+$ polar angle of
 $0.8<\cos\Theta_{K_{CM}}<1.0$.
 The observed differential cross sections were consistent with theoretical
 calculations in order of magnitude.
 However, in view of our limited statistics, further data are needed
 for more quantitative discussions, which will be available in future
 experiments at SPring-8/LEPS or TJNAF.

 We thank the staff at SPring-8 for providing excellent experimental
 conditions during the experiment. 
We thank Dr.~A.~Hosaka, Dr.~D.~Jido and Dr.~S.I.~Nam for helpful discussion.
This work was supported by the Grant-in-Aid
for Scientific Research from the
Ministry of Education, Culture, Science and Technology,
Japan.

\clearpage
 \bibliography{thebibliography}

\end{document}